\documentclass[sigconf]{acmart}

\setcopyright{acmlicensed}
\copyrightyear{2026}
\acmYear{2026}
\setcopyright{cc}
\setcctype{by}
\acmConference[KDD 2026] {Proceedings of the 32nd ACM SIGKDD Conference on Knowledge Discovery and Data Mining V.2}{August 9--13, 2026}{Jeju Island, Republic of Korea.}
\acmBooktitle{Proceedings of the 32nd ACM SIGKDD Conference on Knowledge Discovery and Data Mining V.2 (KDD 2026), August 9--13, 2026, Jeju Island, Republic of Korea}
\acmISBN{979-8-4007-2259-2/2026/08}
\acmDOI{10.1145/3770855.3817630}

\usepackage{multirow}

\usepackage{amsmath,amssymb,amsfonts}
\usepackage{algorithmic}
\usepackage{graphicx}
\usepackage{textcomp}
\usepackage{color, colortbl}
\usepackage{subfigure}
\usepackage{algorithm}
\usepackage{tcolorbox} 
\usepackage{enumitem}
\usepackage{balance}
\usepackage{marvosym}
\makeatletter
\renewcommand{\@authornotemark}{%
  \g@addto@macro\@currentauthors{%
    \advance\hfuzz by 5pt\relax
    \textsuperscript{\Letter}}}

\renewcommand{\authornote}[1]{%
  \if@ACM@anonymous\else
    \g@addto@macro\addresses{\@authornotemark}%
    \g@addto@macro\@authornotes{%
      \begingroup
      \renewcommand{\thefootnote}{\Letter}%
      \stepcounter{footnote}\footnotetext{#1}%
      \endgroup}%
  \fi}
\makeatother

\definecolor{backred}{RGB}{255, 190, 190}
\definecolor{backblue}{RGB}{210, 230, 250}
\definecolor{verylightgray}{gray}{0.95}
\definecolor{skyblue}{RGB}{135, 206, 235}

\newcommand{\ours}{{MixRAGRec}}

\begin{document}

\title[Mixture-of-Experts Knowledge Graph RAG for Multi-Agent LLM-based Recommendation]{Mixture-of-Experts Knowledge Graph Retrieval-Augmented Generation for Multi-Agent LLM-based Recommendation}


\author{Shijie Wang}
\email{shijie.wang@connect.polyu.hk}
\affiliation{%
  \institution{The Hong Kong Polytechnic University}
  \city{Hong Kong SAR}
  \country{China}
}

\author{Chengyi Liu}
\email{chengyi.liu@connect.polyu.hk}
\affiliation{%
  \institution{The Hong Kong Polytechnic University}
  \city{Hong Kong SAR}
  \country{China}
}

\author{Yujuan Ding}\authornote{Corresponding author}
\email{dingyujuan385@gmail.com}
\affiliation{%
  \institution{The Hong Kong Polytechnic University}
  \city{Hong Kong SAR}
  \country{China}
}

\author{Shanru Lin}
\email{lllam32316@gmail.com}
\affiliation{%
  \institution{The Hong Kong Polytechnic University}
  \city{Hong Kong SAR}
  \country{China}
}

\author{See-Kiong Ng}
\email{seekiong@nus.edu.sg}
\affiliation{%
  \institution{National University of Singapore}
  \city{Singapore}
  \country{Singapore}
}

\author{Xu Xin}
\email{xin.xu@polyu.edu.hk}
\affiliation{%
  \institution{The Hong Kong Polytechnic University}
  \city{Hong Kong SAR}
  \country{China}
}

\author{Wenqi Fan}
\email{wenqifan03@gmail.com}
\affiliation{%
  \institution{The Hong Kong Polytechnic University}
  \city{Hong Kong SAR}
  \country{China}
}

\renewcommand{\shortauthors}{Shijie Wang et al.}

\begin{abstract}
Large language models (LLMs) have recently been adopted for recommendations due to their ability to understand user intent and item semantics. 
However, LLM-based recommender systems often rely on parametric knowledge and suffer from outdated knowledge, motivating knowledge graph retrieval-augmented generation (KG-RAG) to ground recommendations on structured, up-to-date KGs.
Despite this promise, effective KG-RAG in recommendations faces great challenges.
First, users' queries vary in complexity and require KG knowledge at different granularities, whereas existing methods adopt a one-size-fits-all retrieval strategy, leading to over-retrieval for simple queries and under-retrieval for complex ones.
In addition, augmenting LLMs with KG knowledge requires translating graph-structured data into linear text, which may introduce noise and cause structural information loss.
Moreover, the selection of retrieval granularity lacks direct supervision and must be inferred from the final recommendation after alignment and downstream utilization, making query-aware retrieval hard to learn end-to-end.
To address these issues, we propose \emph{\ours{}}, a cooperative multi-agent framework for KG-RAG recommendations.
\ours{} integrates a Mixture-of-Experts Retrieval Agent that routes each query to a KG retrieval expert with different granularities, a Knowledge Preference Alignment Agent that converts structured knowledge into LLM-friendly natural language, and a Contrastive Learning-reinforced Recommendation Agent trained with contrastive preference feedback.
Notably, we introduce Mixture-of-Experts Multi-Agent Policy Optimization (MMAPO) to train three agents under a unified objective.
Extensive experiments on real-world datasets demonstrate the effectiveness of our framework.  
Our codes are available at 
\url{https://github.com/Sjay-Wang/MixRAGRec}.

\end{abstract}




\maketitle
\newcommand\kddavailabilityurl{https://doi.org/10.5281/zenodo.20372864}
\ifdefempty{\kddavailabilityurl}{}{
\begingroup\small\noindent\raggedright\textbf{Resource Availability:}\\
The source code of this paper has been made publicly available at \url{\kddavailabilityurl}.
\endgroup
}


\section{Introduction}

\begin{figure}[t]
\centering
\vskip -0.1in
\centering
{\includegraphics[width=0.95\linewidth]{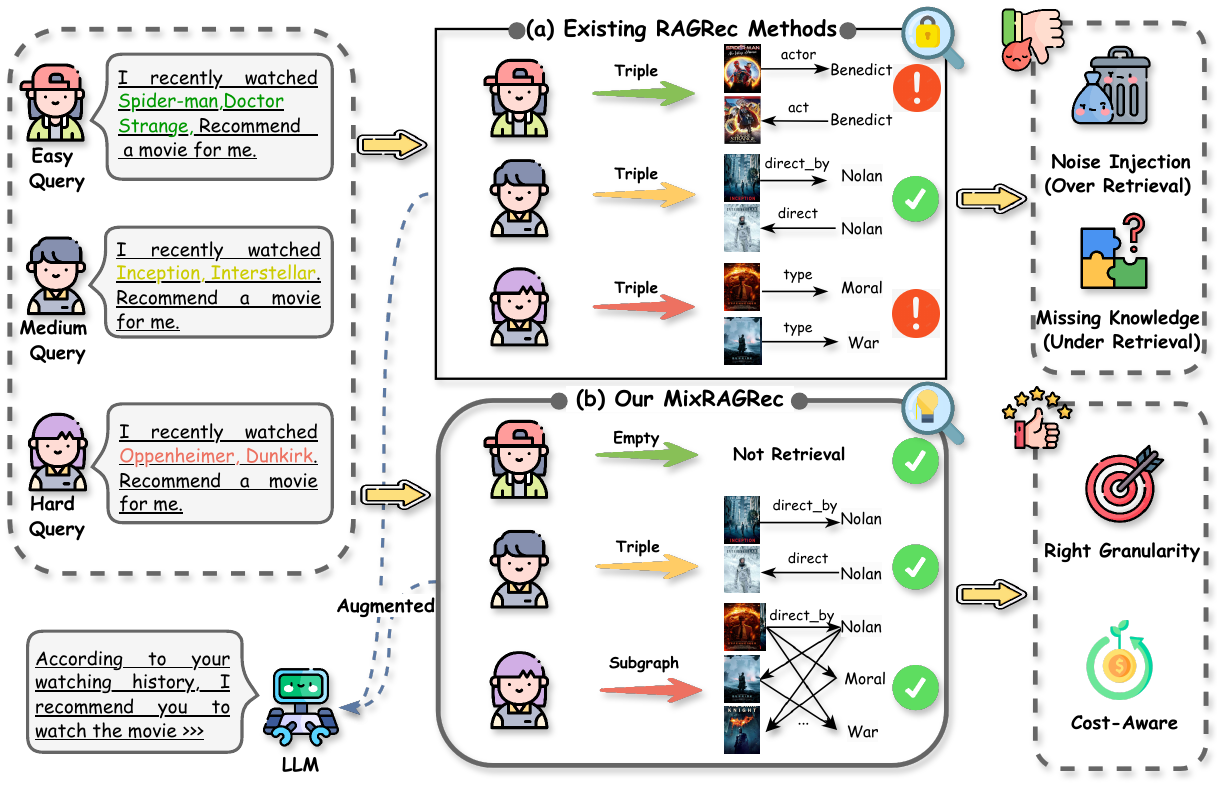}}
\vskip -0.1in
\caption{
Illustration of existing RAGRec methods and our proposed \ours{} methods.
(a) Existing RAGRec methods apply the same KG retrieval granularity to all queries, causing either noise from over-retrieval or missing knowledge from under-retrieval; (b) Our \ours{} routes queries of different complexity to appropriate retrieval granularity, improving the utility–cost trade-off. 
} 
\label{fig:illustration}
\vskip -0.2in
\end{figure}
Recommender systems, as a key technique for alleviating information overload and enabling personalized decision-making, have attracted increasing attention across domains such as e-commerce and social media~\cite{schafer2001commerce, fan2019graph,ding2023modeling,wang2024multi}. 
Recently, Large Language Models (LLMs) have achieved significant breakthroughs and show strong potential as a new paradigm for recommender systems~\cite{bao2023tallrec,ni2022sentence,zhao2024recommender}. By training on vast and diverse data, LLMs exhibit strong generalization and nuanced language understanding capabilities~\cite{ni2026streasoner,xu2026comprehensive,naveed2025comprehensive,chang2024survey}. 
These capabilities enable LLM-based recommender systems to distill user preferences from profiles, item metadata, and interaction histories for more tailored suggestions~\cite{zhang2024notellm,qu2025tokenrec}.

Despite their remarkable advantages in language understanding and generalization, existing LLM-based recommender systems still suffer from critical limitations, including hallucinations and insufficient up-to-date, domain-specific knowledge~\cite{wang2025knowledge,lin2024rella}. 
For instance, an LLM may recommend a non-existent sequel like \textit{``The Matrix 5: Reawakened''} to a fan of the ``Matrix''. 
Furthermore, the static nature of LLMs' training data often leaves them unable to recommend the latest items absent from the pre-training corpus. 
To alleviate these issues, recent Retrieval-Augmented Generation (RAG) leveraging external Knowledge Graph (KG) has emerged as a promising solution to provide the latest and domain-specific structured knowledge without costly fine-tuning~\cite{wang2025knowledge, qiu2025graph}. 
By grounding LLMs in \emph{external, structured, updated}, and \emph{editable} knowledge sources, KG-RAG provides a robust mechanism for supplying factual, up-to-date, and domain-specific information, 
thereby mitigating hallucinations and knowledge latency.

However, achieving effective KG-RAG recommender systems faces significant challenges.
\emph{First}, user queries require KG knowledge at different granularities, while existing KG-RAG recommender systems typically adopt a one-size-fits-all retrieval granularity~\cite{huang2025towards,wang2025knowledge}. This mismatch leads to over-retrieval for simple queries (introducing redundant context and extra cost) and under-retrieval for complex queries (missing multi-hop relations needed for reasoning), as illustrated in Figure~\ref{fig:illustration}(a).
Hence, it remains challenging to design a query-aware retrieval mechanism that selects an appropriate retrieval granularity to balance knowledge utility and retrieval cost.
\emph{Additionally}, the retrieved knowledge is naturally graph-structured, while the input to LLMs is plain text. Directly describing structured knowledge into text can introduce redundant noise and lose structural information, reducing the utility of retrieved knowledge~\cite{perozzi2024let}.
Thus, it is challenging to convert KG knowledge into an LLM-friendly representation that preserves recommendation-relevant knowledge while avoiding unnecessary noise.
\emph{Furthermore}, the selection of retrieval granularity lacks direct supervision and can only be inferred from the final recommendation outcome after alignment and downstream utilization.
This intricate interlinking makes the influence of retrieval granularity hard to disentangle, while applying vanilla end-to-end optimization with only the recommendation criteria may converge to a fixed granularity preference. 
This poses a special challenge to effectively coordinate the retrieval, 
alignment, and recommendation under a unified training objective.

To address these challenges, we propose \textbf{\ours{}} (\textbf{Mix}ture-of-Experts KG-\textbf{RAG} for LLM-based \textbf{Rec}ommendation), a cooperative multi-agent KG-RAG framework that integrates query-aware retrieval, knowledge alignment, and knowledge-augmented recommendation.
Specifically, \ours{} consists of three cooperating agents: 
(1) a \textbf{Mixture-of-Experts Retrieval Agent} that routes each query to one of multiple KG retrieval experts with different granularities, ranging from \emph{no retrieval} (direct generation) to triple retrieval, neighborhood subgraph retrieval, and connected-graph retrieval;
(2) a \textbf{Knowledge Preference Alignment Agent} that transforms retrieved structured knowledge into an LLM-friendly natural-language knowledge snippet, bridging the gap between graph-structured knowledge and textual inputs;
and (3) a \textbf{Contrastive Learning-reinforced Recommendation Agent} trained with \textbf{contrastive preference feedback}, which encourages the model to rank the target item above competitive negatives under the same context.
To jointly optimize these agents, we introduce \textbf{MMAPO} (\textbf{M}ixture-of-Experts \textbf{M}ulti-\textbf{A}gent \textbf{P}olicy \textbf{O}ptimization), which coordinates discrete expert selection, knowledge alignment, and downstream recommendation under a shared objective.
In particular, MMAPO incorporates a \textbf{marginal information gain} term that quantifies retrieval utility relative to its computational cost, encouraging the system to retrieve richer knowledge only when the expected benefit justifies the added overhead. The main contributions of this paper can be summarized as follows:
\begin{itemize}[leftmargin=*]
\item We propose \ours{}, a multi-agent KG-RAG framework for LLM-based recommendation that performs query-aware retrieval with different granularities via a mixture of KG  experts.
\item We introduce a Knowledge Preference Alignment agent that converts structured KG knowledge into concise, LLM-friendly natural language, mitigating the structure-semantics gap.
\item We propose MMAPO, a unified optimization framework that couples agents with a shared objective and a retrieval-utility reward for cost-aware retrieval.
\item We conduct comprehensive experiments on various real-world datasets to demonstrate the effectiveness of the proposed \ours{} framework.
\end{itemize}
\section{Related Work}

\begin{figure*}[htb]
    \centering
    \vskip -0.15in
    \includegraphics[width=0.9\textwidth]{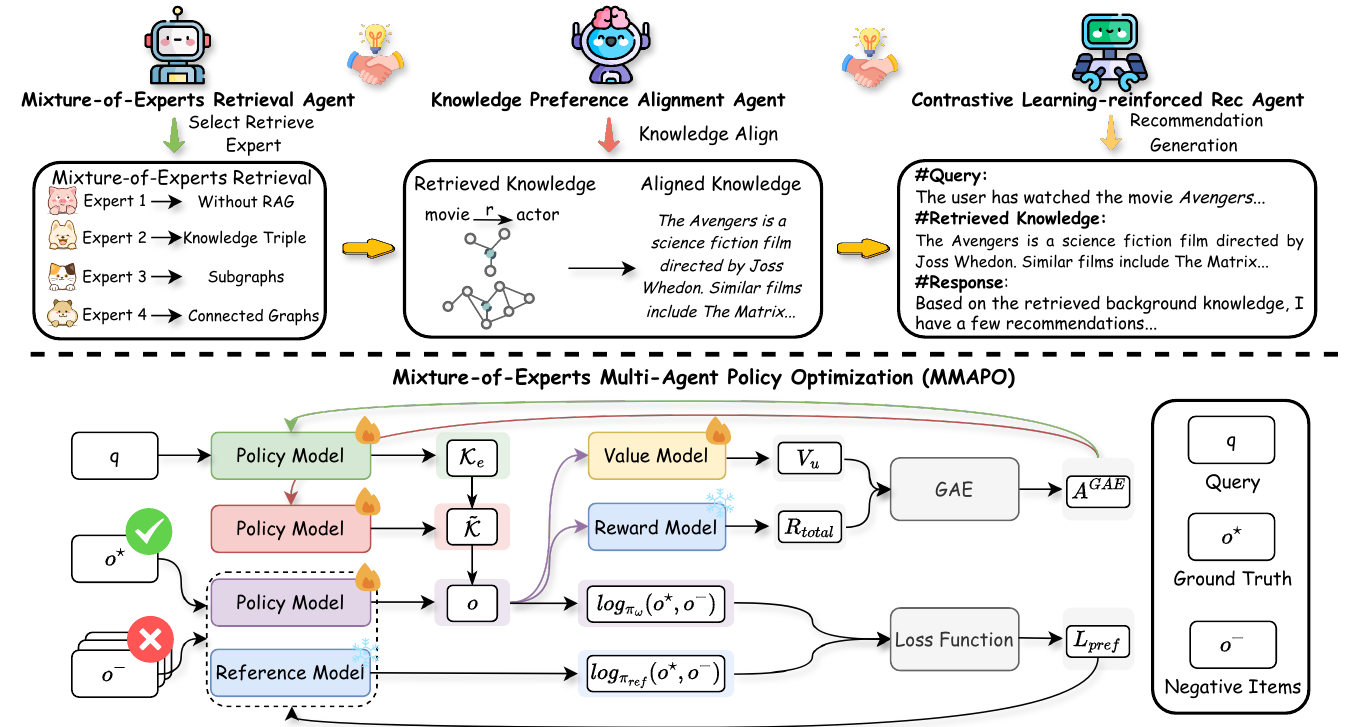}
    \caption{The overview of the \ours{}, which consists of
three agents: (1) \emph{Mixture-of-Experts Retrieval Agent} selects a retrieval expert to obtain structured knowledge, (2) \emph{Knowledge Preference Alignment Agent} converts the retrieved KG knowledge into an LLM-friendly representation, and (3) \emph{Contrastive Learning-reinforced Recommendation Agent} produces the recommendation. \emph{Mixture-of-Experts Multi-Agent Policy Optimization (MMAPO)} jointly optimizes these agents under a shared objective.
    }
    \label{fig:framework}
    \vskip -0.15in
\end{figure*}

Retrieval-Augmented Generation (RAG) has emerged as a robust strategy to mitigate LLM limitations, such as hallucinations and lack of knowledge, by grounding outputs in external knowledge sources~
\cite{fan2024survey,lewis2020retrieval,gao2023retrieval,gumaan2025expertrag}. 
Early studies such as RETRO~\cite{borgeaud2022improving} and FLARE~\cite {jiang2023active} typically retrieve fragments from unstructured corpora (e.g., passages or documents) and condition generation on the retrieved context.
Beyond text corpora, recent studies explore augmenting LLMs with graph-structured knowledge (e.g., KGs) to provide structured knowledge for LLM inference~\cite{he2024g,mavromatis2025gnn,han2024retrieval,wang2025graph}.
Representative studies such as  G-retriever~\cite{he2024g} and GNN-RAG~\cite{mavromatis2025gnn} retrieve subgraphs from KGs and incorporate them into LLM inference via either structured encoders or text-based serialization.

Building on these advances, RAG has been introduced into recommender systems to ground LLM-based recommendations on external knowledge, mitigating issues such as hallucinations and outdated knowledge.
Depending on the source of retrieved knowledge, existing RAG-based recommender systems can be broadly grouped into two types.
The first type of methods typically retrieve knowledge from datasets (e.g., historical user--item interactions and user behaviors) to provide domain-specific knowledge for LLMs to improve user modeling~\cite{yang2025cold,zhu2025collaborative,lin2024rella,xu2025rallrec,li2025g,maragheh2025arag}.
For instance, ReLLa~\cite{lin2024rella} and RALLRec~\cite{xu2025rallrec} retrieve the most relevant behaviors from long interaction histories instead of relying solely on the most recent actions, while G-Refer~\cite{li2025g} retrieves user profiles and explanation paths from interaction graphs to enhance explainability.
The other methods retrieve knowledge from external knowledge bases, most notably knowledge graphs, to provide structured and up-to-date knowledge for LLM-based recommendations~\cite{wang2025knowledge,qiu2025graph,zhao2025webrec}.
A representative method, K-RagRec~\cite{wang2025knowledge}, retrieves relevant knowledge subgraphs for cold-start items to improve knowledge-augmented recommendation.
Despite their success, existing methods are typically built on fixed retrieval and ignore the difficulties of queries, which can lead to suboptimal accuracy and efficiency.
To fill this gap, we propose \ours{} to enable query-aware retrieval and effective utilization of retrieved KG knowledge within a unified framework. We provide more related work in Appendix~\ref{app:related}.

\section{Methodology}

In this section, we first introduce some key notations and concepts in this paper.
Then, we provide the details for each component of our proposed framework \ours{}.

\subsection{Problem Formulation}
In this paper, we consider a knowledge-augmented recommendation scenario. For each instance, the system receives a user query $q$ describing the user intent and profile context (e.g., historical interactions), and outputs a recommended item $o$ from a candidate set $\mathcal{O}$. To support retrieval-augmented recommendation, the system can access a KG $G=(\mathcal{V},\mathcal{R},\mathcal{T})$, where $\mathcal{V}$ is the node set, $\mathcal{R}$ is the relation set, and $\mathcal{T}\subseteq \mathcal{V}\times\mathcal{R}\times\mathcal{V}$ is the triple set. 
Given a query $q$, the system may retrieve a knowledge set from $G$. 
We define our retrieval expert set with four experts, which is extendable to more. Specifically, $\mathcal{E}_{\text{exp}}=\{e_1,e_2,e_3,e_4\}$ and each $e$ returns structured knowledge $\mathcal{K}_{e}$ at a specific granularity.

It is worth noting that \ours{} is formulated as a cooperative multi-agent decision process.
Generally, given $q$,  
a Mixture-of-Experts retrieval agent chooses an expert $e\in\mathcal{E}_{\text{exp}}$ to execute retrieval; a knowledge preference alignment agent transforms the retrieved knowledge $\mathcal{K}_{e}$ into a natural-language representation $\tilde{\mathcal{K}}$; finally, a recommendation agent outputs $o$ conditioned on $(q,\tilde{\mathcal{K}},\mathcal{O})$. 
The framework aims to optimize recommendation performance while accounting for retrieval utility. 
Formally, the objective function is defined as:
\begin{equation}
{\small
\begin{aligned}
&(\theta^\ast,\psi^\ast,\omega^\ast)
=
\arg\max_{\theta,\psi,\omega}\;
\mathbb{E}_{(q_i,\mathcal{O}_i,o_i^\star)\sim\mathcal{D}}
\Big[
\mathbb{E}_{e_i \sim \pi_{\theta}(\cdot \mid s_i)}
\;\\
&\mathbb{E}_{\tilde{\mathcal{K}}_i \sim \pi_{\psi}(\cdot \mid q_i, \mathcal{K}_{i,e_i})}
\;
\mathbb{E}_{o_i \sim \pi_{\omega}(\cdot \mid q_i,\tilde{\mathcal{K}}_i,\mathcal{O}_i)}
\big[
R(q_i,\mathcal{O}_i,o_i^\star;\, e_i,\tilde{\mathcal{K}}_i,o_i)
\big]
\Big],
\end{aligned}
}
\label{eq:overall_objective}
\end{equation}         
\noindent where $\pi_\theta$, $\pi_\psi$, and $\pi_\omega$ are the policies of the retrieval, alignment, and recommendation agents, respectively. $\mathcal{D}$ denotes the training distribution and $o^\star \in \mathcal{O}$ is the ground-truth item. $s_i$ is the retrieval state constructed from the query context, and $R(\cdot)$ is the reward function detailed in Section~\ref{reward}.

\subsection{The Overview of \ours{}}
As illustrated in Figure~\ref{fig:framework}, \ours{} is a multi-agent framework for knowledge-augmented recommendation that consists of three crucial agents: the \emph{Mixture-of-Experts Retrieval Agent}, the \emph{Knowledge Preference Alignment Agent}, and the \emph{Contrastive Learning-reinforced Recommendation Agent}. 
Specifically, given a user query, the Mixture-of-Experts retrieval agent selects the appropriate retrieval expert to obtain structured knowledge from the KG. Next, the knowledge preference alignment agent converts the retrieved knowledge into an LLM-friendly representation, bridging the representational gap between graph-structured knowledge and text-based modeling. Finally, the contrastive learning-reinforced recommendation agent produces the recommendation based on the query and aligned knowledge. 
To train these agents under a shared objective and enable query-aware expert selection, we propose \emph{Mixture-of-Experts Multi-Agent Policy Optimization (MMAPO)}, which coordinates them through a shared reward that accounts for both recommendation performance and retrieval utility. This design enables \ours{} to generate accurate recommendations by adapting retrieval for diverse queries through a unified multi-agent framework.

\subsection{Multi-Granularity KG Retrieval with Mixture-of-Experts}
\ours{} adopts a Mixture-of-Experts retrieval module that retrieves structured knowledge from an item-centric KG at multiple granularities.
As illustrated in Figure~\ref{fig:retrieve}, the left part shows KG construction and indexing, while the right part depicts a Mixture-of-Experts retrieval design, ranging from direct generation without retrieval to triple-, subgraph-, and connected-graph retrieval. This design aims to provide query-relevant knowledge while avoiding unnecessary retrieval cost and noise.

\subsubsection{KG Construction and Indexing}
We represent the knowledge source as a KG $G=(\mathcal{V},\mathcal{R},\mathcal{T})$, where $\mathcal{V}$ and $\mathcal{R}$ denote the sets of entities and relations, and $\mathcal{T}\subseteq \mathcal{V}\times\mathcal{R}\times\mathcal{V}$ denotes the set of triples $(h,r,t)$. 
To facilitate flexible retrieval, we encode both entities and triples into a shared embedding space using a pre-trained language encoder. For an entity $n\in\mathcal{V}$ with textual description $x_n$, we capture the semantic information as $\mathbf{z}_n = \mathrm{Encode}(x_n)\in \mathbb{R}^d,$ where $d$ denotes the dimension of the output representation.
Then we build an entity vector database $\mathcal{Z}_{\mathcal{V}}=\{\mathbf{z}_n\}_{n\in\mathcal{V}}$. 
Similarly, for a triple $(h,r,t)\in\mathcal{T}$ with text descriptions $(x_h,x_r,x_t)$, we compute
\begin{equation}
\mathbf{z}_{(h,r,t)}=\mathrm{Encode}\big(\mathrm{Concat}(x_h, x_r, x_t)\big)\in \mathbb{R}^d,
\end{equation}
and construct a triple vector database $\mathcal{Z}_{\mathcal{T}}=\{\mathbf{z}_{(h,r,t)}\}_{(h,r,t)\in\mathcal{T}}$.
By utilizing this dual indexing, retrieval experts can access KG knowledge at different levels of granularity.

\subsubsection{Mixture-of-Experts KG Retrieval}
While existing KG-RAG methods provide a general mechanism to incorporate KG knowledge, they typically rely on a fixed retrieval strategy. In practice, user queries vary widely, ranging from simple cases requiring little or no external knowledge to complex ones involving multi-hop knowledge subgraphs. As a result, a fixed KG retrieval strategy may under-retrieve for complex queries, leading to insufficient knowledge for accurate recommendation, or over-retrieve for simple queries, introducing unnecessary noise and computational overhead.

\begin{figure}[t]
    \centering
    \includegraphics[width=0.9\linewidth]{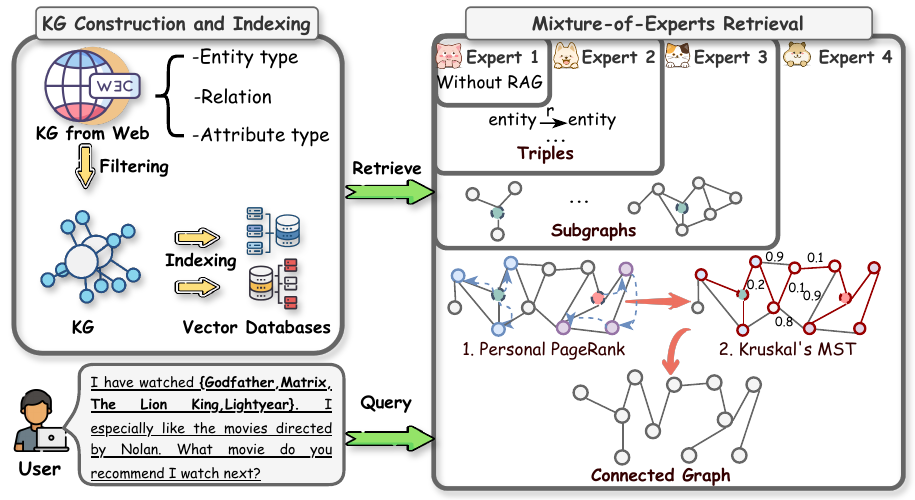}
    \vskip -0.1in
    \caption{Demonstration of KG Construction and Indexing and the Retrieval Experts.  }
    \label{fig:retrieve}
    \vskip -0.2in
\end{figure}

To this end, we propose a Mixture-of-Experts retrieval agent that selects the appropriate retrieval granularity for each query. 
Specifically, four retrieval experts are designed to address diverse query scenarios, including direct generation without retrieval, triple-level retrieval for precise facts, $k$-hop neighborhood subgraphs capturing higher-order relations, and connected subgraphs representing global knowledge, as shown in Figure~\ref{fig:retrieve}.

Given a user query $q$, we compute its embedding using the same encoder as in indexing to ensure a consistent representation space: $\mathbf{z}_q=\mathrm{Encode}(q)\in\mathbb{R}^d.$

The Mixture-of-Experts retrieval agent then selects one expert $e\in\{1,2,3,4\}$ and returns the retrieved KG knowledge $\mathcal{K}_e$, as illustrated in Figure~\ref{fig:retrieve} (right).

\noindent\textbf{Expert 1: DirectGenerator (no retrieval).}
This expert performs no KG retrieval and returns an empty set, i.e., $\mathcal{K}_1=\varnothing$. 

\noindent\textbf{Expert 2: TripleRetriever (triple-level retrieval).}
This expert retrieves the most relevant KG triples by semantic similarity.
Formally, the retrieved triple set is:
\begin{equation}
\mathcal{K}_2 = \operatorname{TopM}_{(h,r,t)\in\mathcal{T}}
\ \mathrm{sim}\big(\mathbf{z}_q,\mathbf{z}_{(h,r,t)}\big),
\end{equation}
where $M$ is the retrieval budget, and is instantiated differently for different granularities. $\mathrm{sim}(\cdot,\cdot)$ measures the cosine similarity.

\noindent\textbf{Expert 3: SubgraphRetriever ($k$-hop neighborhood subgraph).}
To capture multi-hop relational context, this expert first selects a seed set of relevant entities:
\begin{equation}
\mathcal{S} = \operatorname{TopM}_{n\in\mathcal{V}}
\ \mathrm{sim}\big(\mathbf{z}_q,\mathbf{z}_n\big),
\end{equation}
and then expands each seed into its $k$-hop neighborhood subgraph.
Denote the $k$-hop neighborhood subgraph around entity $n$ as $G_n^{(k)}$, the retrieved knowledge is the union $\mathcal{K}_3 = \bigcup_{n\in\mathcal{S}} G_n^{(k)}$.

\noindent\textbf{Expert 4: ConnectedGraphRetriever (connected subgraph).}
For complex queries that benefit from globally coherent knowledge, the expert constructs a connected subgraph that captures global knowledge. It begins with the same seed set $\mathcal{S}$ as Expert 3, and then executes a Personalized PageRank (PPR)~\cite{page1999pagerank} from $\mathcal{S}$ to obtain importance scores over entities. 
Then the top-$M$ entities by PPR score, denoted $\mathcal{V}_{\mathrm{PPR}}$, will be selected and consider the induced subgraph $G[\mathcal{V}_{\mathrm{PPR}}]$. 
To obtain a connected subgraph containing the most relevant entities while minimizing edge costs, we formulate the retrieval objective as finding a minimum-cost spanning tree (MST) over $G[\mathcal{V}_{\mathrm{PPR}}]$. 
Specifically, we assign each edge $e$ a cost based on the semantic match between the query and the relation type of the edge as $\mathrm{cost}(e)=1-\cos\!\left(\mathbf{z}_q,\mathbf{z}_{r(e)}\right),$
where $\mathbf{z}_q$ is the query embedding, and
$\mathbf{z}_{r(e)}=\mathrm{Encode}(x_{r(e)})$ is the embedding of the relation type associated with edge $e$  computed by the same language encoder used in KG indexing,
with $x_{r(e)}$ being the textual description of relation $r(e)$.
Then we compute the MST by applying Kruskal's algorithm~\cite{kruskal1956shortest}:
\begin{equation}
\mathcal{K}_4 = \mathrm{MST}\big(G[\mathcal{V}_{\mathrm{PPR}}]\big).
\end{equation}

Kruskal's algorithm greedily adds edges in increasing order of $\mathrm{cost}(\cdot)$ while avoiding cycles, thereby connecting the subgraph with important entities via relations most relevant to the query.

\subsection{Knowledge Preference Alignment}
While the Mixture-of-Experts retrieval agent provides structured KG knowledge, the retrieved triples or subgraphs are misaligned with the natural-language inputs expected by LLM-based recommenders. Naive textual serialization can introduce noise and lose structural information, degrading recommendation performance. 
To address this issue, \ours{} introduces a knowledge preference alignment agent that converts retrieved KG knowledge into LLM-friendly representations while preserving recommendation-relevant information.

Let $\mathcal{K}$ denote the retrieved knowledge returned by the selected retrieval expert. Expert~2 returns a set of triples, Expert~3 returns a $k$-hop neighborhood subgraph, Expert~4 returns a connected subgraph, and Expert~1 corresponds to no retrieval. 
Since the $\mathcal{K}$ is structured, we first transform it into a stable textual draft using a carefully crafted template module $\mathrm{Temp}(\cdot)$. To be more specific, for triple-level retrieval (i.e., Expert~2), $\mathrm{Temp}(\cdot)$ converts each triple $(h,r,t)$ into a short natural-language statement with relation-aware patterns, e.g.,
\emph{``\{head\} has \{relation\} \{tail\}''},
and concatenates them into a compact list of knowledge statements.
For graph-based retrieval (i.e., Expert~3 and Expert~4), $\mathrm{Temp}(\cdot)$ linearizes the retrieved graph into a set of salient relational statements (e.g., edges along short paths among the retrieved entities), providing a consistent input format across retrieval granularities.

Given the query $q$ and the templated text $\mathrm{Temp}(\mathcal{K})$, the knowledge preference alignment agent further refines the draft into an aligned knowledge snippet
$\tilde{\mathcal{K}} = \mathrm{Align}_{\psi}\!\left(q, \mathrm{Temp}(\mathcal{K})\right),$
where $\mathrm{Align}_{\psi}(\cdot)$ is the knowledge preference alignment agent and trained to produce knowledge that is maximally useful to the downstream recommendation agent. The aligned knowledge $\tilde{\mathcal{K}}$ serves as a textual interface between structured KG knowledge and LLM-based recommendations and is appended to the downstream recommendation prompt along with the user's query.

\subsection{Preference-Optimized Recommendation via Contrastive Feedback}\label{sec:pref_opt_rec}
In knowledge-augmented recommendation, users' contexts often contain numerous reasonable items that are highly similar to the target item, such as items sharing the same genre, entities, or overlapping attributes. Such similar items can easily confuse the model, especially when external knowledge from the KG is integrated. Directly optimizing the recommender may fail to sufficiently differentiate the target items from these strong alternatives, reducing the recommendation performance.
To this end, we introduce a contrastive learning-reinforced recommendation agent to optimize the recommendation agent with contrastive preference feedback.

Specifically, given a user's query $q$ and aligned knowledge $\tilde{\mathcal{K}}$, we define the recommendation context as $x=(q,\tilde{\mathcal{K}})$. Let $\mathcal{O}$ denote the candidate item set under context $x$, $o^\star\in\mathcal{O}$ is defined as the target item and other items $o^{-}\in \mathcal{O}\setminus\{o^\star\}$ are treated as negative items. The objective is to train the model to assign a higher probability to the positive item than to negative items. 
However, directly optimizing this objective is costly and less efficient due to the large scale of the action space. To solve this problem, we introduce hard negative sampling to select the top-$N$  competitive negative items according to the current recommendation policy, encouraging the model to learn sharper distinctions among semantically similar items. Formally, we define the hard negative set as:
\begin{equation}
\mathcal{N}_{\mathrm{hard}}(x)
=
\operatorname{TopN}_{o \in \mathcal{O}\setminus\{o^\star\}}
\ \pi_{\omega}(o \mid x),
\label{eq:hard_neg}
\end{equation}
and construct preference pairs $(x, o^\star, o^{-})$ with $o^{-}\in \mathcal{N}_{\mathrm{hard}}(x)$. The training objective is then defined as:
{\setlength{\abovedisplayskip}{4pt}
 \setlength{\belowdisplayskip}{4pt}
 \setlength{\abovedisplayshortskip}{2pt}
 \setlength{\belowdisplayshortskip}{2pt}

\begin{equation}
\mathcal{L}_{\mathrm{pref}}
=
-\log \sigma\!\left(
\beta \left[
\log \frac{\pi_{\omega}(o^\star\mid x)}{\pi_{\mathrm{ref}}(o^\star\mid x)}
-
\log \frac{\pi_{\omega}(o^{-}\mid x)}{\pi_{\mathrm{ref}}(o^{-}\mid x)}
\right]
\right),
\label{eq:pref_obj}
\end{equation}}

\noindent where $\sigma(\cdot)$ is the sigmoid function and $\beta$ is a temperature hyperparameter.
We introduce $\pi_{\mathrm{ref}}$ as an anchoring baseline to stabilize learning, and optimize $\pi_{\omega}$ to assign higher probability to the target item. When multiple hard negatives are available, we compute Eq.~\eqref{eq:pref_obj} for each $(o^\star,o^{-})$ pair and average the loss across negatives.

\subsection{Mixture-of-Experts Multi-Agent Policy Optimization (MMAPO)}\label{optimization}
Due to the intricate interconnections among the three agents, it is not trivial to train ~\ours{}. Specifically, the recommendation performance depends on the selected retrieval expert, knowledge alignment quality, and the recommendation agent's capabilities.
These tight interconnections make it difficult to evaluate each agent's influence during training.
Therefore, vanilla optimization methods make the model tend to select the expert based on fixed preferences rather than selecting experts based on the query.

To this end,  we propose Mixture-of-Experts Multi-Agent Policy Optimization (MMAPO), which coordinates the three agents under a shared objective and enables the model to identify the most suitable retrieval expert based on the user's query, as shown in Figure~\ref{fig:framework}. Specifically, MMAPO couples agents via a shared reward that combines recommendation performance with a marginal information gain. This design encourages flexible expert selection that balances knowledge richness and efficiency, while jointly improving knowledge alignment and recommendation performance.

\subsubsection{Advantage estimation}
In \ours{}, the retrieval expert selected by the Mixture-of-Experts retrieval agent affects the final reward only through the downstream aligned knowledge and the subsequent recommendation.
This delayed dependency makes the retrieval policy difficult to optimize from raw returns.
Therefore, MMAPO adopts advantage-based credit assignment with learned value baselines.

We unfold the pipeline into a trajectory of decision steps, where the retrieval action and the alignment token generations are treated as actions at different steps. Let $r_t$ denote the reward at step $t$ and define the discounted return
$G_t = \sum_{l=0}^{T-t}\gamma^l r_{t+l},$
where $\gamma\in(0,1]$ is the discount factor. We use a sparse terminal reward $r_t=0$ for $t<T$ and $r_T=R_{\mathrm{total}}$.
MMAPO maintains value baselines for the retrieval and alignment agents.
Specifically, the retrieval agent uses a value network $V_\phi(s_t)$ over its state $s_t$, and the alignment agent is equipped with a value head $V_\psi(c_t)$ defined on its generation context $c_t$ (i.e., the prompt plus the generated prefix).
Let $V(\cdot)$ denote the corresponding value baseline (i.e., $V(u_t)=V_\phi(s_t)$ for retrieval steps, and $V(u_t)=V_\psi(c_t)$ for alignment steps).
We compute the temporal-difference residual:
$\delta_t = r_t + \gamma V(u_{t+1}) - V(u_t),$
and estimate advantages using Generalized Advantage Estimation (GAE):
\begin{equation}
A_t^{\mathrm{GAE}} = \sum_{l=0}^{T-t}(\gamma\lambda)^l\,\delta_{t+l},
\label{eq:gae}
\end{equation}
where $\lambda\in[0,1]$ controls the bias--variance trade-off.
This estimator provides low-variance learning signals and propagates downstream reward back to the retrieval decision and the alignment generation process through their respective value baselines.

\subsubsection{Reward Design}\label{reward}
To encourage flexible expert selection that balances recommendation performance and retrieval utility, we introduce a two-dimensional reward function, as follows:
\begin{equation}
R_{\mathrm{total}} = R_{\mathrm{rec}} + \alpha\cdot R_{\mathrm{MIG}},
\label{eq:rtotal}
\end{equation}
where $\alpha$ is the weighting coefficient for coordinating the two reward dimensions. The recommendation reward $R_{\mathrm{rec}}$ measures the performance of the final recommendation:
\begin{equation}
R_{\mathrm{rec}} =
\begin{cases}
\mathrm{conf}(o), & \text{if $o=o^\star$},\\
-0.1, & \text{otherwise},
\end{cases}
\label{eq:rrec}
\end{equation}
\noindent where $\mathrm{conf}(o)=\pi_\omega(o\mid q,\tilde{\mathcal{K}},\mathcal{O})$ denotes the model confidence assigned to the recommended item.

To make expert selection cost-aware, we introduce a marginal information gain reward:
\begin{equation}
R_{\mathrm{MIG}}=\Delta I(e)-\eta\cdot C(e), \
\text{where} \ \Delta I(e)=D_{\mathrm{KL}}\!\left(P_{\mathrm{expert}} \,\|\, P_{\mathrm{base}}\right),
\label{eq:rmig}
\end{equation}
where $P_{\mathrm{expert}}$ denotes the recommendation distribution when expert $e$ is invoked,
$P_{\mathrm{base}}$ is a baseline distribution (e.g., Expert~1 without retrieval),
and $C(e)$ is a computational cost assigned to expert $e$, increasing with retrieval complexity.
By rewarding information gain while penalizing retrieval cost, $R_{\mathrm{MIG}}$ encourages expensive retrieval only when it yields sufficient utility to justify the added cost.
MMAPO optimizes the retrieval and alignment policies under this shared reward while training the recommendation agent with contrastive preference feedback.
The core procedure is summarized in Algorithm~\ref{alg:mmapo_core}, and the full training objective and update rules are provided in Appendix~\ref{app:mmapo_updates}.

\section{Experiments}

In this section, we evaluate the effectiveness of \ours{} through extensive experiments. 
We will first introduce the experimental settings, including the datasets, compared baseline methods, evaluation metrics, and implementation details. 
Then, we report the main experimental results, highlighting the performance of \ours{} compared with various strong baselines.
Finally, we conduct comprehensive additional analyses, including efficiency evaluation, ablation studies, and hyper-parameter analysis.

\begin{table}[]
 \vskip -0.02in
  \centering
  \caption{Basic statistics of three datasets and the KGs. ``Items in KG'' indicates the number of items that appeared in both the KG and the dataset. 
  }
   \vskip -0.1in
    \scalebox{0.6}
{
  \resizebox{0.7\textwidth}{!}{
    \begin{tabular}{cccc}
    \toprule
    \multirow{1}{*}{\textbf{Datasets}} & {\textbf{MovieLens-1M}} & {\textbf{MovieLens-20M}} & {\textbf{LFM-1K}} \\
    \midrule
\textbf{User}  & 6,040 & 138,287 & 992 \\
\textbf{Item} & 3,883 & 27,278 & 107,528 \\
\textbf{Interaction} & 1,000,209 & 20,000,263 & 19,150,868 \\
\textbf{Items in KG} & 3,253 & 24,700 & 33,736 \\
\midrule
\textbf{Entities} & 24,840 & 72,673 & 180,457\\
\textbf{Relations} &  168   & 223    & 312 \\
\textbf{KG triples} & 64,688 & 272,173 & 516,604  \\
    \bottomrule
    \end{tabular}%
    }
}
 \vskip -0.1in
  \label{tab:datasets}%
\end{table}%

\begin{table*}[]
\centering
\caption{Performance comparison across LLM-based recommendation baselines and KG-RAG methods. Best results are in \textbf{bold} and the second-best results are \underline{underlined}.}
  \vskip -0.1in
\resizebox{0.9\textwidth}{!}{
\begin{tabular}{lccccccccc}
\toprule      
 \multirow{2}{*}{\textbf{Model}}     & \multicolumn{3}{c}{\textbf{MovieLens-1M}} & \multicolumn{3}{c}{\textbf{MovieLens-20M }} & \multicolumn{3}{c}{\textbf{LFM-1K}} \\                                            & Accuracy   & Recall@3   & Recall@5  & Accuracy    & Recall@3    & Recall@5    & Accuracy    & Recall@3    & Recall@5                 \\ \midrule
\multicolumn{10}{c}{\cellcolor{gray!20}\textit{\textbf{Zero-shot Inference}}} \\
Gemini-2.0-Flash                                      &0.389     &N/A  &N/A  &0.467   &N/A & N/A  &0.728  & N/A  &N/A \\
GPT-4o                                 &0.420   &N/A  &N/A  &0.524  &N/A & N/A  &0.685 & N/A  &N/A \\
Deepseek-r1                                & 0.393  &N/A  &N/A  &0.445  &N/A  &N/A   & 0.634 &N/A  &N/A \\
LLaMA3-8B\textsubscript{zero-shot}                             & 0.130  &N/A  &N/A  &0.129  &N/A & N/A  &0.234 & N/A  &N/A \\
\multicolumn{10}{c}{\cellcolor{gray!20}\textit{\textbf{LLM-based Recommendation Models}}} \\
TallRec\textsubscript{LLaMA3-8B}~\cite{bao2023tallrec}                                         &0.391   &\underline{0.663}   &\underline{0.779} &0.521  &\underline{0.781}  &\underline{0.868} &0.699 &0.848   &0.891                   \\
Rec-r1\textsubscript{LLaMA3-8B}~\cite{lin2025rec}                              &   0.402&  0.558       &0.634   &\underline{0.594}  &0.761  &0.815  &\underline{0.872} &\underline{0.926}   &\underline{0.938}              \\
\multicolumn{10}{c}{\cellcolor{gray!20}\textit{\textbf{KG-RAG for LLM-based Recommendation Models}}} \\
KG-Text\textsubscript{LLaMA3-8B}~\cite{wu2023retrieve}                         &0.234   & N/A    & N/A  &0.301  &N/A  &N/A  &0.545 & N/A  &  N/A                  \\
KAPING\textsubscript{LLaMA3-8B}~\cite{baek2023knowledge}                                  & 0.232  &N/A   &N/A  &0.274   &N/A  & N/A &0.530  &N/A & N/A                       \\
G-retriever\textsubscript{LLaMA3-8B}~\cite{he2024g}                        &0.395    & 0.648   & 0.743  & 0.491  & 0.722 & 0.801 & 0.690  & 0.825  & 0.860      \\
K-RagRec\textsubscript{LLaMA3-8B}~\cite{wang2025knowledge}                      & \underline{0.454}  &0.659  & 0.707  &0.578  &0.720  &0.755  &0.845 &0.886   &0.906   \\
\rowcolor{red!15} 
    \textbf{\ours{}\textsubscript{LLaMA3-8B}}\textbf{(Ours)}                               & \textbf{0.504}  & \textbf{0.798}  & \textbf{0.882}  & \textbf{0.676}   & \textbf{0.900} & \textbf{0.940}  & \textbf{0.934}   & \textbf{0.990}  & \textbf{0.992} \\
\rowcolor{blue!15} Improvement                          & 11.0\% & 20.4\% & 13.2\% & 13.8\% & 15.2\% & 8.3\% & 7.1\% & 6.9\%  & 5.8\%       \\
\multicolumn{10}{c}{\cellcolor{gray!20}\textit{\textbf{Other Backbone LLM}}} \\
Mistral-7B\textsubscript{zero-shot}                                 &0.105   &N/A  &N/A  &  0.097 &N/A & N/A  &0.318   & N/A  &N/A \\
TallRec\textsubscript{Mistral-7B}~\cite{bao2023tallrec}                                         &0.393  &0.665  &0.761 &0.545 &\underline{0.809}  &0.883  &0.674 &0.812   &0.867                   \\
Rec-r1\textsubscript{Mistral-7B}~\cite{lin2025rec}                                &0.338   &0.433         &0.490   &0.505  &0.599  &0.648  &0.362 &0.477   &0.552              \\
\toprule 
KG-Text\textsubscript{Mistral-7B}~\cite{wu2023retrieve}                        &0.228   &N/A  &N/A  &  0.277 &N/A & N/A  &0.492   & N/A  &N/A \\
KAPING\textsubscript{Mistral-7B}~\cite{baek2023knowledge}                                  &0.249   & N/A    & N/A  &0.272  &N/A  &N/A  &0.536 & N/A  &  N/A    \\
G-retriever\textsubscript{Mistral-7B}~\cite{he2024g}                       &0.372  & 0.641 &0.775 & 0.542  & 0.758   & 0.846    &0.763 &0.856   &0.890       \\
K-RagRec\textsubscript{Mistral-7B}~\cite{wang2025knowledge}            & \underline{0.438}  &  \underline{0.727}  &  \underline{0.831} & \underline{0.573} & 0.801   &\underline{0.888}     &\underline{0.818}  & \underline{0.874} & \underline{0.894}     \\
\rowcolor{red!15} 
\textbf{\ours{}\textsubscript{Mistral-7B}}\textbf{(Ours)}                               &  \textbf{0.518} & \textbf{0.785}  & \textbf{0.879} & \textbf{0.652} & \textbf{0.892 } & \textbf{0.935}  & \textbf{0.912}   & \textbf{0.953}   &  \textbf{0.962}
\\ 
\rowcolor{blue!15} Improvement                         &18.3\%  &8.0\%  &5.8\%  & 13.8\% & 10.3\% & 5.3\% & 11.5\% & 9.0\%  & 7.6\%      \\
\bottomrule
\end{tabular}
}
\label{tab:results}
\end{table*}

\subsection{Datasets}
We evaluate the performance of \ours{} on three widely used real-world recommendation datasets covering both movie and music domains. 
\textbf{MovieLens-1M}\footnote{https://grouplens.org/datasets/movielens/1m/} contains approximately one million user-movie ratings, along with basic textual metadata such as movie titles. 
\textbf{MovieLens-20M}\footnote{https://grouplens.org/datasets/movielens/20m/} is a large-scale movie dataset containing over 20 million ratings from more than 138,000 users for approximately 27,000 movies. 
\textbf{LastFM-1K}\footnote{http://ocelma.net/MusicRecommendationDataset/lastfm-1K.html} is a music dataset that records more than 19 million user listening histories and associated artist information.
Following our framework design, we adopt DBpedia\footnote{https://www.dbpedia.org/} as the data source and construct corresponding knowledge graphs for each dataset.
The statistics of the datasets and their corresponding KGs are summarized in Table~\ref{tab:datasets}.

\subsection{Baselines}
We compare \ours{} with a carefully constructed set of baselines to provide a comprehensive evaluation.
These baselines cover (i) Zero-shot Inference, (ii)LLM-based Recommendation Models, and (iii) KG-RAG for LLM-based Recommendation Models.

\noindent \textbf{Zero-shot inference.}
These baselines directly prompt LLMs such as \textit{Gemini-2.0-Flash}, \textit{GPT-4o}~\cite{hurst2024gpt}, \textit{Deepseek-r1}~\cite{guo2025deepseek}, \textit{LLaMA3-8B}~\cite{dubey2024llama}, and \textit{Mistral-7B} to perform recommendations without any task-specific fine-tuning or external knowledge augmentation.

\noindent\textbf{LLM-based Recommendation Models.}
These methods adapt LLMs to recommendation tasks by fine-tuning on recommendation datasets.
We include \textit{TallRec}~\cite{bao2023tallrec}, which fine-tunes LLMs using instruction-tuning for recommendation, and \textit{Rec-r1}~\cite{lin2025rec}, which further adapts LLMs for reasoning-augmented recommendation by reinforcement learning.

\noindent\textbf{KG-RAG for LLM-based Recommendation Models.}
These baseline models augment LLMs by retrieving structured knowledge from the KG and are adapted in our experiments to support recommendation tasks.
We include triple-level retrieval methods KG-Text~\cite{wu2023retrieve} and subgraph-level retrieval methods KAPING~\cite{baek2023knowledge} and G-retriever~\cite{he2024g}. We also compare \ours{} with K-RagRec~\cite{wang2025knowledge}, designed for KG-RAG recommendation tasks, which retrieves knowledge subgraphs of different hops from KG for knowledge-augmented recommendation.

\subsection{Evaluation Metrics}
We evaluate recommendation performance using Accuracy and Recall@$K$ ($K = 3,5$). 
Accuracy measures whether the model's recommendation matches the target item, while 
Recall@$K$ measures whether the target item appears in the top-$K$ recommended list.
Following recent studies~\cite{zhang2024agentcf,wang2025knowledge}, we select the item most recently interacted with by the user as the target item and use the 10 interactions preceding it as the user context. The model is then required to predict the user's preferred item from a candidate pool consisting of the target item and 19 distractor items. For inference-based models that only output a single option without providing comparable scores over all candidates, we report Accuracy only.

\subsection{Parameter Settings}
Our framework is implemented in PyTorch, and all experiments are conducted on one NVIDIA H20-96G GPU.
For the Mixture-of-Experts retrieval agent and knowledge preference alignment agent, we optimize the policy using a learning rate $3\times 10^{-4}$, discount factor $\gamma=0.99$, GAE parameter $\lambda=0.95$, clipping coefficient $\epsilon=0.2$, and $4$ update epochs per iteration.
For preference optimization in the recommendation agent, we set the temperature parameter to $\beta=0.2$ and employ hard negative sampling with $N=10$ negatives.
For the shared reward design, we set the MIG weight $\alpha=0.2$ and the cost penalty $\eta=0.005$. Notably, the knowledge preference alignment agent and the recommendation agent adopt the same LLM backbone architecture (e.g., LLaMA3-8B and Mistral-7B) with independent parameters. Additional implementation details appear in Appendix~\ref{app:implementation}.

\subsection{Overall Performance Comparison}
We compare the recommendation performance of \ours{} with various carefully constructed baselines under the LLM backbones LLaMA3-8B and Mistral-7B, on three datasets across three metrics. The experimental results are reported in the Table~\ref{tab:results}. From the comparison, we have the following key observations:
\begin{itemize}[leftmargin=*, topsep=2pt, itemsep=1pt, parsep=0pt, partopsep=0pt]
\item Zero-shot prompting is insufficient for recommendation. Across three datasets, zero-shot inference consistently underperforms methods adapted to recommendation (e.g., TallRec/Rec-r1) and knowledge-augmented approaches. Although closed-source LLMs (e.g., GPT-4o and Deepseek-r1) achieved stronger zero-shot Accuracy than open-source backbones (e.g., LLaMA3-8B and Mistral-7B), they still lag considerably behind the best baselines and \ours{}, highlighting the limitation of direct prompting for recommendations.

\item Retrieving external knowledge from the KG can substantially improve recommendation performance, while the benefit depends on how knowledge is retrieved and integrated. 
Directly injecting retrieved KG knowledge as text (e.g., KG-Text and KAPING) does not get consistent improvements over strong LLM-based recommender systems.
In contrast, methods that leverage graph-level retrieval and representation (e.g., G-Retriever and K-RagRec) generally achieve stronger performance, suggesting that preserving structural information is important for effective knowledge augmentation.

\item \ours{} outperforms the strongest baselines on all three datasets under both backbones, with relative improvements ranging from 5.8\% to 20.4\% under LLaMA3-8B and around 5.3\% to 18.3\% under Mistral-7B, highlighting the effectiveness of \ours{}. Notably, \ours{} also demonstrates significant improvement in the Recall@$K$ metric, indicating that the proposed framework improves the quality of the top-$K$ ranked recommendations rather than only the Accuracy.
\end{itemize}

\subsection{Efficiency Study}
To evaluate the efficiency of \ours{}, we measure the average retrieval time and end-to-end total latency on MovieLens-1M with the LLaMA3-8B backbone, using one NVIDIA H20-96G GPU. The results are summarized in Table~\ref{tab:time}.
While lightweight retrieval methods like KG-Text and KAPING present fast retrieval efficiency, they have limited accuracy. 
Although graph-level retrieval methods (e.g., G-retriever and K-RagRec) achieve higher accuracy, they incur significantly higher retrieval time and total latency. 
In particular, K-RagRec exhibits the largest latency because it requires individual retrieval for each item.
In contrast, \ours{} achieves the best accuracy while maintaining significantly lower retrieval time and total latency than baselines, suggesting that flexible expert selection can preserve the benefits of structured knowledge without unnecessary retrieval costs.

We further analyze different retrieval experts in \ours{}. As the retrieval granularity increases from Expert 2 to Expert 4, both retrieval time and accuracy increase, reflecting the expected cost--utility trade-off of richer KG knowledge.
It is worth noting that \ours{} achieves the highest accuracy while keeping retrieval times close to the Expert 2, indicating that the learned policy invokes expensive retrieval only when it provides sufficient utility.
These results highlight that \ours{} improves recommendation performance with lower cost, validating the effectiveness and efficiency of the proposed framework. We also analyze the retrieval complexity in Appendix~\ref{app:complexity}.

\begin{table}[t]
  \centering
  \caption{ 
   Efficiency comparison on MovieLens-1M with the LLaMA3-8B backbone. We report average retrieval time and end-to-end total time in seconds (s). ACC denotes Accuracy.}
  \vskip -0.1in
    \scalebox{0.75}
{
    \begin{tabular}{cccc}
    \toprule
    \multirow{1}{*}{\textbf{Methods}}  & \multirow{1}{*}{\textbf{ACC}} & \multirow{1}{*}{\textbf{Retrieval Time (s)}} & \multirow{1}{*}{\textbf{Total Time (s)}}  \\
    \midrule
    KG-Text  &0.234  & 0.311   &2.974 \\
    KAPING &0.232 & 0.247 &2.708  \\
    G-retriever &0.395  &2.411  &2.907 \\
    K-RagRec & 0.454 &2.867    &3.776 \\
    \midrule
    \rowcolor{skyblue!16} Expert 1 (DirectGenerator) & - & 0.000 &  0.042 \\
    \rowcolor{skyblue!16} Expert 2 (TripleRetriever) & 0.488  & 0.066  & 1.827 \\
    \rowcolor{skyblue!16} Expert 3 (SubgraphRetriever) & 0.491  & 0.212  & 2.103 \\
    \rowcolor{skyblue!16} Expert 4 (ConnectedGraphRetriever) & 0.499 & 0.374  & 2.271 \\
    \rowcolor{blue!8} \textbf{\ours{}} & \textbf{0.504}  & \textbf{0.063}   & \textbf{1.563}\\
    \bottomrule
    \end{tabular}%
}
  \label{tab:time}%
  \vskip -0.18in
\end{table}%

\begin{figure}[t]
\vskip -0.15in
\centering
\subfigure{\includegraphics[width=0.49\columnwidth]{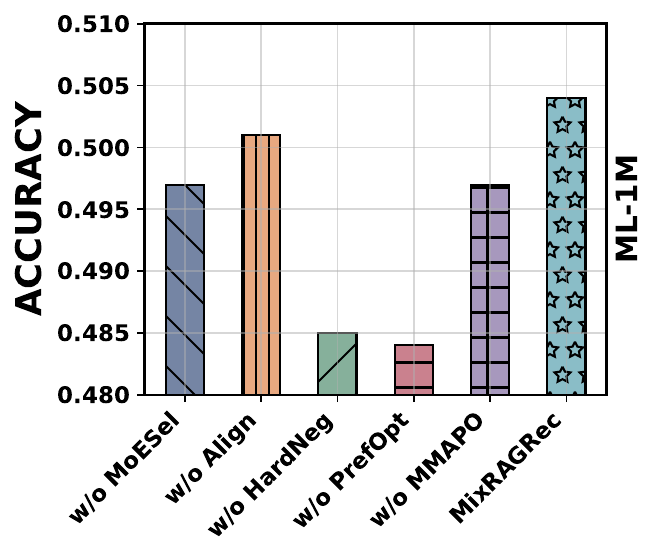}}
\subfigure{\includegraphics[width=0.49\columnwidth]{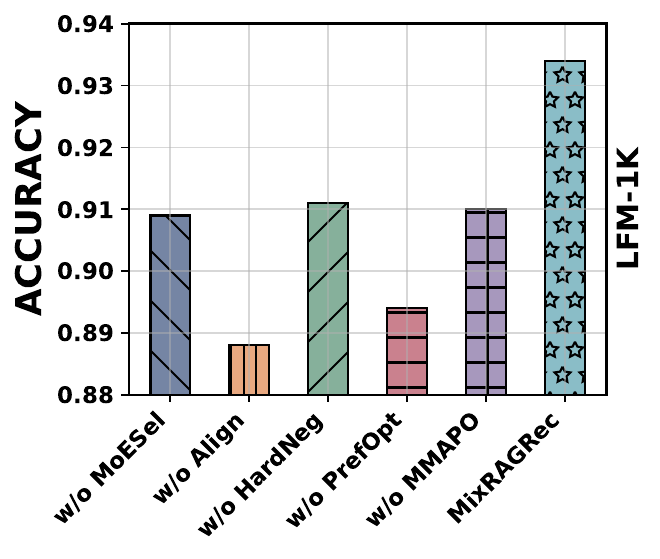}}
\vskip -0.2in
\subfigure{\includegraphics[width=0.49\columnwidth]{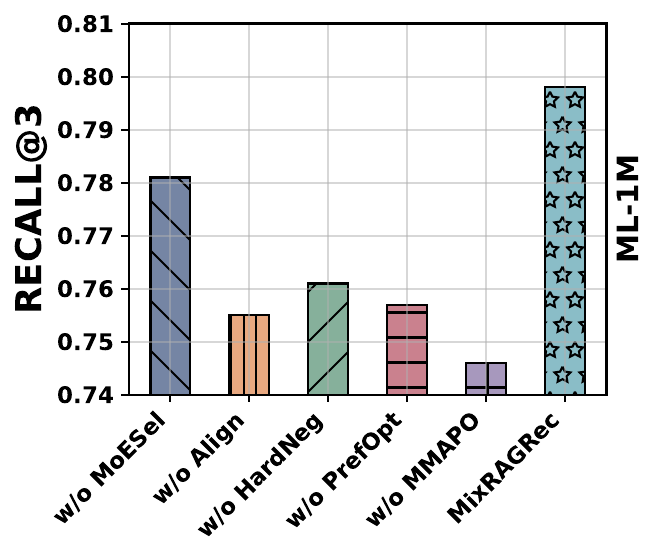}}
\subfigure{\includegraphics[width=0.49\columnwidth]{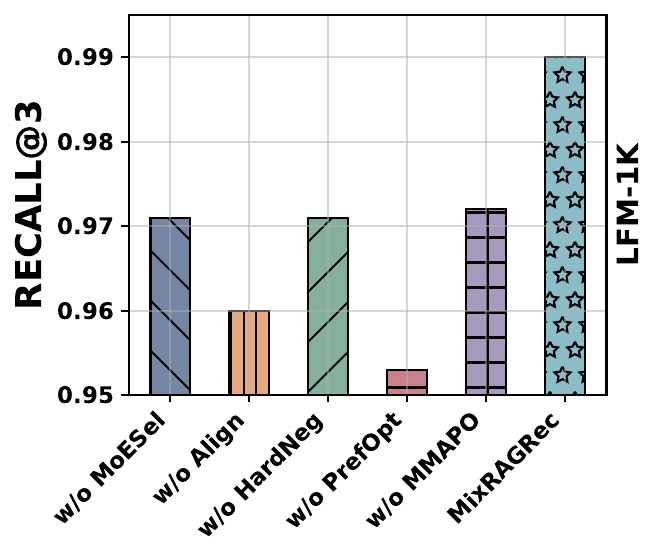}}
\vskip -0.13in
\subfigure{\includegraphics[width=0.49\columnwidth]{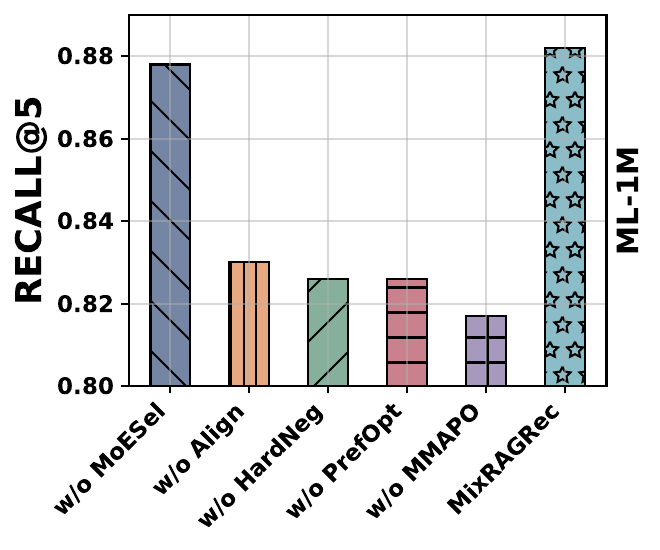}}
\subfigure{\includegraphics[width=0.49\columnwidth]{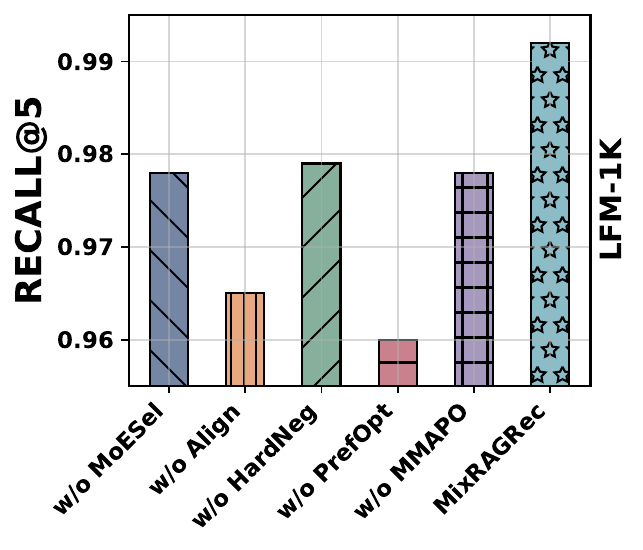}}
\vskip -0.13in
\caption{Ablation results of \ours{} and five variants on MovieLens-1M and LFM-1K with the LLaMA3-8B backbone. We report Accuracy, Recall@3, and Recall@5.}\label{fig:ablation}
\vskip -0.25in
\end{figure}

\subsection{Ablation Study}
We conduct ablation studies with LLaMA3-8B on MovieLens-1M and LFM-1K to assess the contribution of each key component in \ours{}. Specifically, we evaluate five ablated variants: (i) \textbf{w/o MoESel}, which replaces the Mixture-of-Experts selector policy with random expert selection; (ii) \textbf{w/o Align}, which removes the Knowledge Preference Alignment agent and uses only the templated verbalization $\mathrm{Temp}(K)$ as input to the recommender; (iii) \textbf{w/o HardNeg}, which replaces hard negative sampling with random negatives in contrastive preference training; (iv) \textbf{w/o PrefOpt}, which removes contrastive preference feedback for training the recommender; and (v) \textbf{w/o MMAPO},
which removes the shared objective and coordinated updates by dropping the MIG term and optimizing each agent separately with an accuracy-only reward using generic policy-gradient updates. The results are demonstrated in Figure~\ref{fig:ablation}.

Overall, the full \ours{} achieves the best performance across all metrics on both datasets, demonstrating the effectiveness of each component. 
Notably, the w/o Align variant resulted in the most significant performance degradation, especially on the LFM-1K dataset, suggesting that the Knowledge Preference Alignment agent is crucial for bridging the gap between retrieval structured knowledge and LLM inputs.
In addition, removing hard negative sampling or contrastive feedback consistently degrades Accuracy and Recall@$K$ on both datasets, highlighting the importance of explicitly separating the target item from highly competitive negatives in knowledge-augmented recommendation.
Moreover, replacing the Mixture-of-Experts retrieval agent with random expert selection and removing MMAPO both lead to sub-optimal results, demonstrating the effectiveness of our proposed Mixture-of-Experts retrieval and optimization approach.

\subsection{Hyper-parameter Analysis}
In this sub-section, we study the impact of three key hyper-parameters of \ours{} on MovieLens-1M with LLaMA3-8B. Specifically, we evaluate the marginal information gain weight $\alpha$, the number of hard negatives $N$, and the retrieval budget $M$. 

\begin{figure}[t]
\centering
\subfigure[ML1M ACC]{\includegraphics[width=0.32\columnwidth]{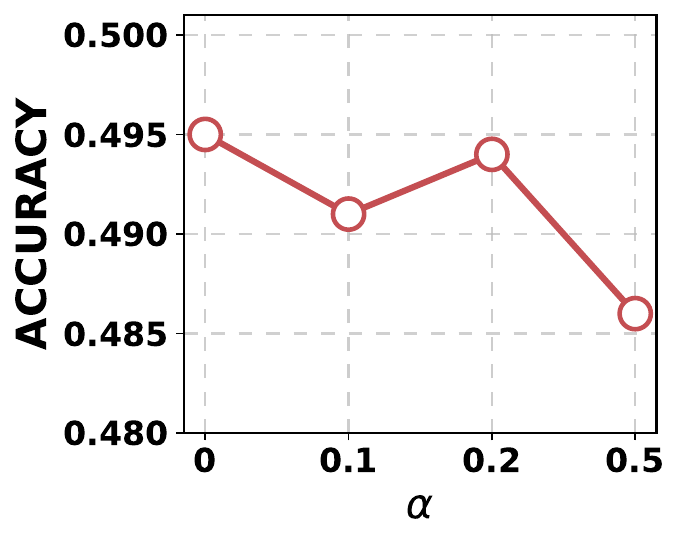}}
\subfigure[ML1M R@3]{\includegraphics[width=0.32\columnwidth]{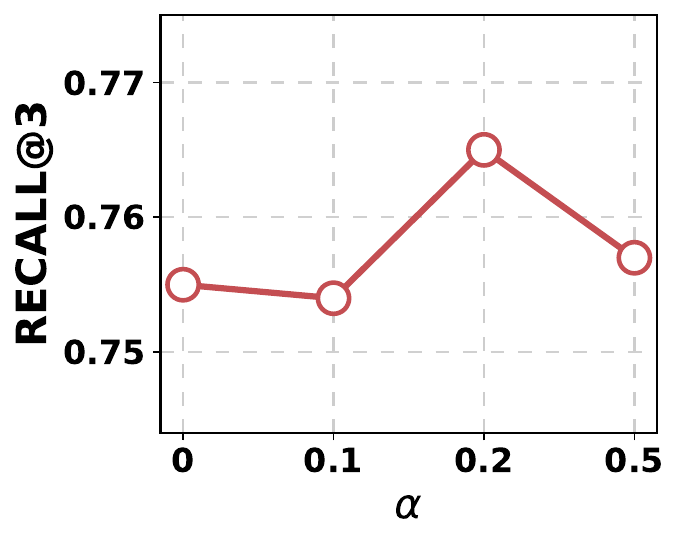}}
\subfigure[Average Retrieval Time]{\includegraphics[width=0.32\columnwidth]{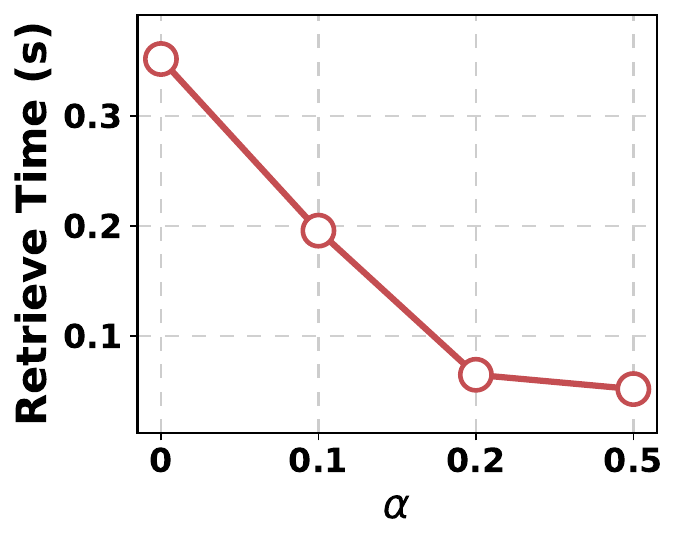}}
\vskip -0.13in
\caption{Effect of the MIG weight $\alpha$ on MovieLens-1M with the LLaMA3-8B backbone.}\label{fig:hyper-mig}
\vskip -0.25in
\end{figure}

\begin{figure}[t]
\centering
\subfigure[ML1M ACC]{\includegraphics[width=0.32\columnwidth]{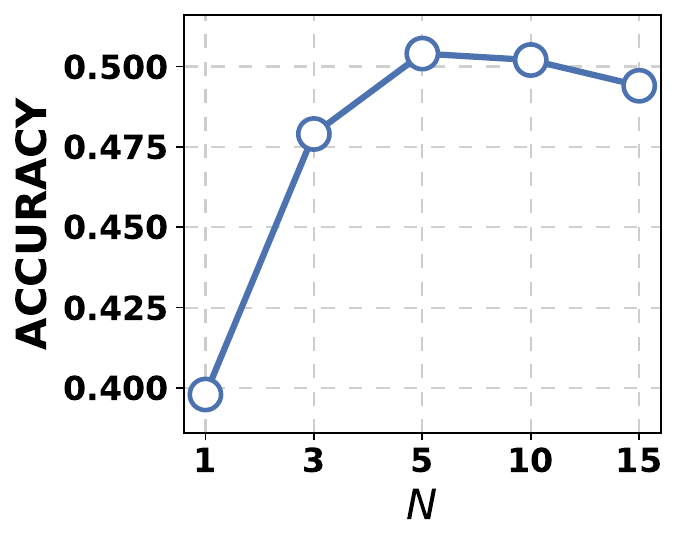}}
\subfigure[ML1M R@3]{\includegraphics[width=0.32\columnwidth]{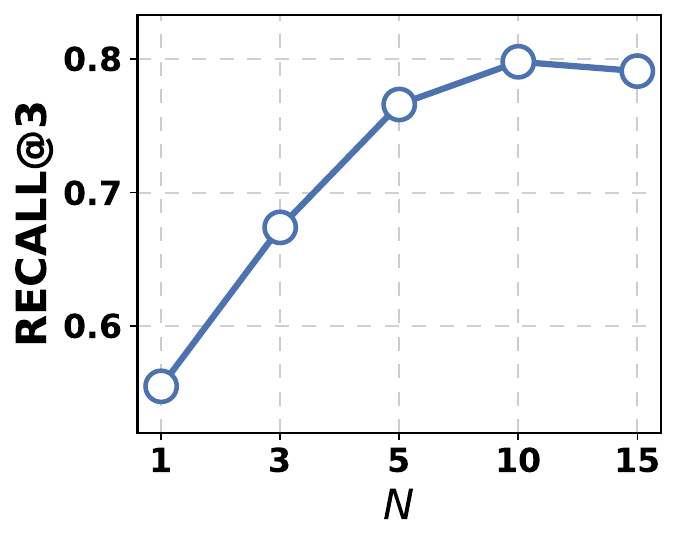}}
\subfigure[ML1M R@5]{\includegraphics[width=0.32\columnwidth]{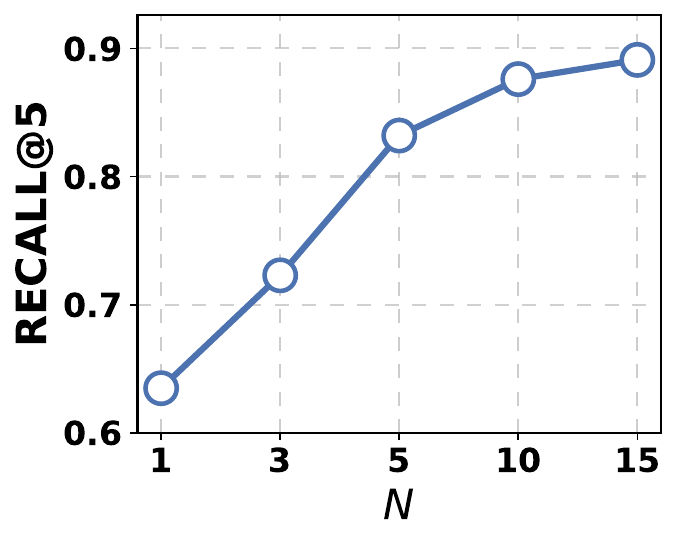}}
\vskip -0.13in
\caption{Effect of the number of hard negatives $N$ on MovieLens-1M with the LLaMA3-8B backbone.}\label{fig:hyper-hard-neg}
\vskip -0.25in
\end{figure}

\begin{figure}[t]
\centering
\subfigure[ML1M ACC]{\includegraphics[width=0.32\columnwidth]{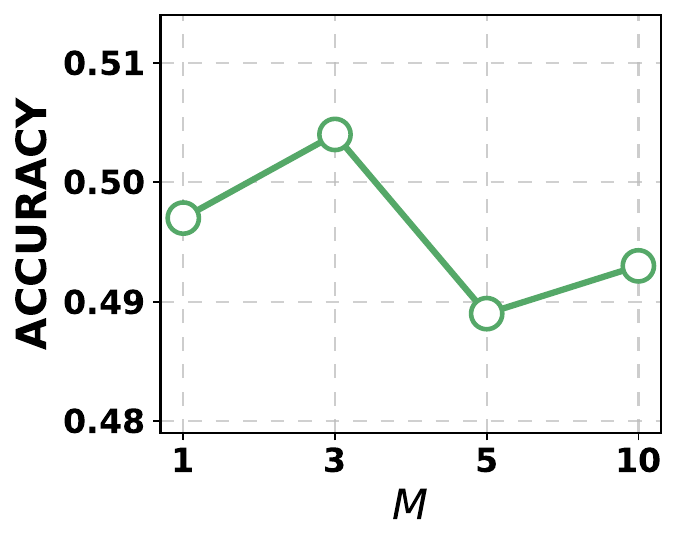}}
\subfigure[ML1M R@3]{\includegraphics[width=0.32\columnwidth]{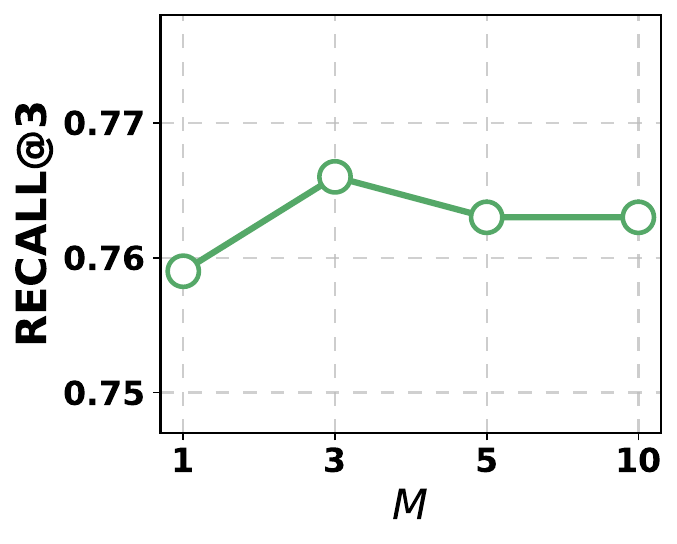}}
\subfigure[ML1M R@5]{\includegraphics[width=0.32\columnwidth]{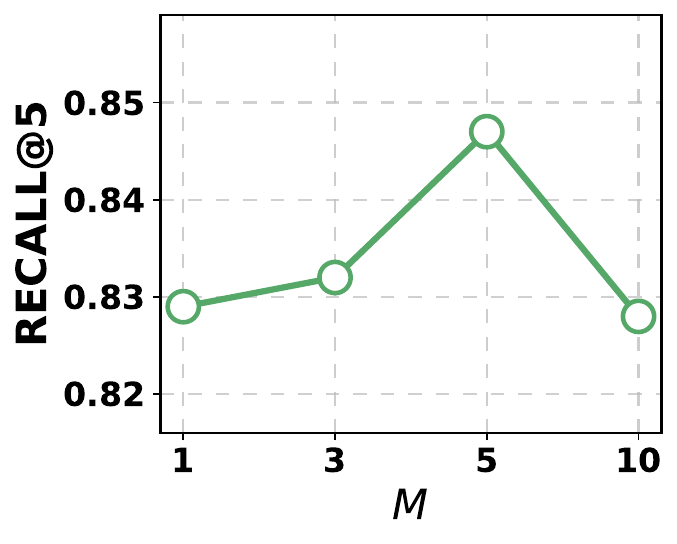}}
\vskip -0.13in
\caption{Effect of the retrieval budget $M$ on MovieLens-1M with the LLaMA3-8B backbone.}\label{fig:hyper-topk}
\vskip -0.2in
\end{figure}

\subsubsection{Effect of the MIG weight $\alpha$.}
We vary the MIG weight $\alpha$ to control the trade-off between retrieval utility and recommendation effectiveness in the shared reward. 
As shown in Figure~\ref{fig:hyper-mig}, setting $\alpha=0$ bases the shared reward solely on recommendation accuracy, resulting in the longest average retrieval time and relatively poorer ranking quality (Recall@3/5). 
Increasing $\alpha$ to a moderate value improves ranking performance while reducing retrieval time. 
In particular, $\alpha=0.2$ achieves the best Recall@3/5 performance while significantly reducing the average retrieval time, suggesting that an appropriate MIG $\alpha$ encourages more cost-aware retrieval while guaranteeing the recommendation performance.
When $\alpha$ is further increased to 0.5, retrieval efficiency improves slightly but performance is significantly reduced.
Overall, these results highlight the importance of selecting an appropriate MIG weight to balance retrieval utility and recommendation effectiveness.

\subsubsection{Effect of the number of hard negatives $N$.}
We analyze the impact of the number of hard negatives on recommendation performance on MovieLens-1M with the LLaMA3-8B backbone.
Results are shown in Figure~\ref{fig:hyper-hard-neg}.
Using too few hard negatives severely degrades performance. For example, when $N=1$, both Accuracy and Recall@3/5 drop substantially, indicating that a single negative provides an insufficient contrastive signal.
Increasing $N$ to 5 and 10 improves both Accuracy and Recall@$K$, suggesting that additional hard negatives help the model better distinguish target items from competitive candidates. 
Further increasing $N$ to 15 yields limited gains with trade-offs across metrics. 
Accordingly, we set $N=10$ in our experiments.

\subsubsection{Effect of the retrieval budget $M$.}
We study the effect of the retrieval budget $M$ on MovieLens-1M with the LLaMA3-8B backbone.
The budget $M$ controls the retrieval scale across experts, and is instantiated differently for different granularities.
For triple-level retrieval, we return the top-$cM$ triples (with $c=5$), while for subgraph-based retrieval, we use $M$ to determine the number of retained nodes (or seed entities) before graph expansion.
Results are shown in Figure~\ref{fig:hyper-topk}.
We observe that a small budget can miss useful knowledge, while a large budget may introduce noise and increase the burden on downstream alignment and utilization.
On MovieLens-1M, $M=3$ and $M=5$ yield optimal Accuracy and Recall, respectively. 
Overall, it is important to select an appropriate retrieval budget based on the scale of the dataset and the KG.

\section{Conclusion}
In this work, we present \ours{}, a cooperative multi-agent framework for KG-RAG recommendation that improves both effectiveness and efficiency by enabling query-aware retrieval and knowledge utilization.
Specifically, \ours{} includes a Mixture-of-Experts Retrieval Agent that selects among KG retrieval experts with different granularities, a Knowledge Preference Alignment Agent that converts structured KG knowledge into LLM-friendly text, and a Contrastive Learning-reinforced Recommendation Agent trained with contrastive preference feedback to better distinguish the target item from competitive candidates.
To jointly train the three agents, we propose Mixture-of-Experts Multi-Agent Policy Optimization (MMAPO), which coordinates learning under a shared objective with recommendation feedback and marginal information gain.
Experiments on three real-world datasets demonstrate the effectiveness and efficiency of \ours{}.

\section*{Acknowledgments}
The research described in this paper has been partially supported by the Key Special Project of National Natural Science Foundation of China (Project No. 72442017), the General Research Funds from the Hong Kong Research Grants Council (project No. PolyU 15207322, 15200023, 15206024, and 15224524), Internal research funds from Hong Kong Polytechnic University (project no. P0059586, P0042693, P0048625, and P0051361), and Sheertek International (HK) Limited. This work was supported by computational resources provided by The Centre for Large AI Models (CLAIM) of The Hong Kong Polytechnic University.

\newpage
\bibliographystyle{ACM-Reference-Format}

\bibliography{sample-base}

@inproceedings{wang2025knowledge,
  title={Knowledge graph retrieval-augmented generation for llm-based recommendation},
  author={Wang, Shijie and Fan, Wenqi and Feng, Yue and Shanru, Lin and Ma, Xinyu and Wang, Shuaiqiang and Yin, Dawei},
  booktitle={Proceedings of the 63rd Annual Meeting of the Association for Computational Linguistics (Volume 1: Long Papers)},
  pages={27152--27168},
  year={2025}
}

@inproceedings{bao2023tallrec,
  title={Tallrec: An effective and efficient tuning framework to align large language model with recommendation},
  author={Bao, Keqin and Zhang, Jizhi and Zhang, Yang and Wang, Wenjie and Feng, Fuli and He, Xiangnan},
  booktitle={Proc. 17th ACM Conf. Recomm. Syst.},
  pages={1007--1014},
  year={2023}
}

@inproceedings{ni2026streasoner,
title={Streasoner: Empowering {LLMs} for Spatio-Temporal Reasoning in Time Series via Spatial-Aware Reinforcement Learning},
author={Ni, Juntong and Wang, Shiyu and He, Qi and Jin, Ming and Jin, Wei},
booktitle={Proceedings of the 64th Annual Meeting of the Association for Computational Linguistics},
year={2026}
}

@article{xu2026comprehensive,
  title={A comprehensive survey of AI Agents in Healthcare},
  author={Xu, Gelei and Li, Xueyang and Chen, Yixiong and Duan, Yuying and Wu, Shuqing and Yu, Haoxinran and Chiu, Ching-Hao and Ni, Juntong and Tang, Ningzhi and Li, Toby Jia-Jun and others},
  journal={Journal of Biomedical Informatics},
  pages={105045},
  year={2026},
  publisher={Elsevier}
}

@article{lin2025rec,
  title={Rec-r1: Bridging generative large language models and user-centric recommendation systems via reinforcement learning},
  author={Lin, Jiacheng and Wang, Tian and Qian, Kun},
  journal={arXiv preprint arXiv:2503.24289},
  year={2025}
}

@article{wu2023retrieve,
  title={Retrieve-rewrite-answer: A kg-to-text enhanced llms framework for knowledge graph question answering},
  author={Wu, Yike and Hu, Nan and Bi, Sheng and Qi, Guilin and Ren, Jie and Xie, Anhuan and Song, Wei},
  journal={arXiv preprint arXiv:2309.11206},
  year={2023}
}

@article{zhao2025webrec,
  title={WebRec: Enhancing LLM-based Recommendations with Attention-guided RAG from Web},
  author={Zhao, Zihuai and Ding, Yujuan and Fan, Wenqi and Li, Qing},
  journal={arXiv preprint arXiv:2511.14182},
  year={2025}
}

@inproceedings{baek2023knowledge,
  title={Knowledge-Augmented Language Model Prompting for Zero-Shot Knowledge Graph Question Answering},
  author={Baek, Jinheon and Aji, Alham Fikri and Saffari, Amir},
  booktitle={Proceedings of the 1st Workshop on Natural Language Reasoning and Structured Explanations (NLRSE)},
  pages={78--106},
  year={2023}
}

@article{he2024g,
  title={G-retriever: Retrieval-augmented generation for textual graph understanding and question answering},
  author={He, Xiaoxin and Tian, Yijun and Sun, Yifei and Chawla, Nitesh and Laurent, Thomas and LeCun, Yann and Bresson, Xavier and Hooi, Bryan},
  journal={Proc. Adv. Neural Inf. Process. Syst.},
  volume={37},
  pages={132876--132907},
  year={2024}
}

@inproceedings{fan2019graph,
  title={Graph neural networks for social recommendation},
  author={Fan, Wenqi and Ma, Yao and Li, Qing and He, Yuan and Zhao, Eric and Tang, Jiliang and Yin, Dawei},
  booktitle={Proc. World Wide Web Conf.},
  pages={417--426},
  year={2019}
}

@article{schafer2001commerce,
  title={E-commerce recommendation applications},
  author={Schafer, J Ben and Konstan, Joseph A and Riedl, John},
  journal={Data mining and knowledge discovery},
  volume={5},
  number={1},
  pages={115--153},
  year={2001},
  publisher={Springer}
}

@inproceedings{ni2022sentence,
  title={Sentence-t5: Scalable sentence encoders from pre-trained text-to-text models},
  author={Ni, Jianmo and Abrego, Gustavo Hernandez and Constant, Noah and Ma, Ji and Hall, Keith and Cer, Daniel and Yang, Yinfei},
  booktitle={Findings of the association for computational linguistics: ACL 2022},
  pages={1864--1874},
  year={2022}
}

@article{qu2025tokenrec,
  title={Tokenrec: Learning to tokenize id for llm-based generative recommendations},
  author={Qu, Haohao and Fan, Wenqi and Zhao, Zihuai and Li, Qing},
  journal={IEEE Trans. Knowl. Data Eng.},
  year={2025},
  publisher={IEEE}
}

@inproceedings{lin2024rella,
  title={Rella: Retrieval-enhanced large language models for lifelong sequential behavior comprehension in recommendation},
  author={Lin, Jianghao and Shan, Rong and Zhu, Chenxu and Du, Kounianhua and Chen, Bo and Quan, Shigang and Tang, Ruiming and Yu, Yong and Zhang, Weinan},
  booktitle={Proc. World Wide Web Conf.},
  pages={3497--3508},
  year={2024}
}

@inproceedings{qiu2025graph,
  title={Graph Retrieval-Augmented LLM for Conversational Recommendation Systems},
  author={Qiu, Zhangchi and Luo, Linhao and Zhao, Zicheng and Pan, Shirui and Liew, Alan Wee-Chung},
  booktitle={Pacific-Asia Conference on Knowledge Discovery and Data Mining},
  pages={344--355},
  year={2025},
  organization={Springer}
}

@article{naveed2025comprehensive,
  title={A comprehensive overview of large language models},
  author={Naveed, Humza and Khan, Asad Ullah and Qiu, Shi and Saqib, Muhammad and Anwar, Saeed and Usman, Muhammad and Akhtar, Naveed and Barnes, Nick and Mian, Ajmal},
  journal={ACM Transactions on Intelligent Systems and Technology},
  volume={16},
  number={5},
  pages={1--72},
  year={2025},
  publisher={ACM New York, NY}
}

@article{chang2024survey,
  title={A survey on evaluation of large language models},
  author={Chang, Yupeng and Wang, Xu and Wang, Jindong and Wu, Yuan and Yang, Linyi and Zhu, Kaijie and Chen, Hao and Yi, Xiaoyuan and Wang, Cunxiang and Wang, Yidong and others},
  journal={ACM transactions on intelligent systems and technology},
  volume={15},
  number={3},
  pages={1--45},
  year={2024},
  publisher={ACM New York, NY}
}

@inproceedings{zhang2024notellm,
  title={Notellm: A retrievable large language model for note recommendation},
  author={Zhang, Chao and Wu, Shiwei and Zhang, Haoxin and Xu, Tong and Gao, Yan and Hu, Yao and Chen, Enhong},
  booktitle={Companion Proceedings of the ACM Web Conference 2024},
  pages={170--179},
  year={2024}
}

@article{zhao2024recommender,
  title={Recommender systems in the era of large language models (llms)},
  author={Zhao, Zihuai and Fan, Wenqi and Li, Jiatong and Liu, Yunqing and Mei, Xiaowei and Wang, Yiqi and Wen, Zhen and Wang, Fei and Zhao, Xiangyu and Tang, Jiliang and others},
  journal={IEEE Trans. Knowl. Data Eng.},
  volume={36},
  number={11},
  pages={6889--6907},
  year={2024},
  publisher={IEEE}
}

@article{perozzi2024let,
  title={Let your graph do the talking: Encoding structured data for llms},
  author={Perozzi, Bryan and Fatemi, Bahare and Zelle, Dustin and Tsitsulin, Anton and Kazemi, Mehran and Al-Rfou, Rami and Halcrow, Jonathan},
  journal={arXiv preprint arXiv:2402.05862},
  year={2024}
}

@article{hurst2024gpt,
  title={Gpt-4o system card},
  author={Hurst, Aaron and Lerer, Adam and Goucher, Adam P and Perelman, Adam and Ramesh, Aditya and Clark, Aidan and Ostrow, AJ and Welihinda, Akila and Hayes, Alan and Radford, Alec and others},
  journal={arXiv preprint arXiv:2410.21276},
  year={2024}
}

@article{guo2025deepseek,
  title={Deepseek-r1: Incentivizing reasoning capability in llms via reinforcement learning},
  author={Guo, Daya and Yang, Dejian and Zhang, Haowei and Song, Junxiao and Zhang, Ruoyu and Xu, Runxin and Zhu, Qihao and Ma, Shirong and Wang, Peiyi and Bi, Xiao and others},
  journal={arXiv preprint arXiv:2501.12948},
  year={2025}
}

@article{dubey2024llama,
  title={The llama 3 herd of models},
  author={Dubey, Abhimanyu and Jauhri, Abhinav and Pandey, Abhinav and Kadian, Abhishek and Al-Dahle, Ahmad and Letman, Aiesha and Mathur, Akhil and Schelten, Alan and Yang, Amy and Fan, Angela and others},
  journal={arXiv e-prints},
  pages={arXiv--2407},
  year={2024}
}

@inproceedings{zhang2024agentcf,
  title={Agentcf: Collaborative learning with autonomous language agents for recommender systems},
  author={Zhang, Junjie and Hou, Yupeng and Xie, Ruobing and Sun, Wenqi and McAuley, Julian and Zhao, Wayne Xin and Lin, Leyu and Wen, Ji-Rong},
  booktitle={Proc. World Wide Web Conf.},
  pages={3679--3689},
  year={2024}
}

@inproceedings{geng2022recommendation,
  title={Recommendation as language processing (rlp): A unified pretrain, personalized prompt \& predict paradigm (p5)},
  author={Geng, Shijie and Liu, Shuchang and Fu, Zuohui and Ge, Yingqiang and Zhang, Yongfeng},
  booktitle={Proc. 16th ACM Conf. Recomm. Syst.},
  pages={299--315},
  year={2022}
}

@article{zhang2025collm,
  title={Collm: Integrating collaborative embeddings into large language models for recommendation},
  author={Zhang, Yang and Feng, Fuli and Zhang, Jizhi and Bao, Keqin and Wang, Qifan and He, Xiangnan},
  journal={IEEE Trans. Knowl. Data Eng.},
  year={2025},
  publisher={IEEE}
}

@inproceedings{fan2024survey,
  title={A survey on rag meeting llms: Towards retrieval-augmented large language models},
  author={Fan, Wenqi and Ding, Yujuan and Ning, Liangbo and Wang, Shijie and Li, Hengyun and Yin, Dawei and Chua, Tat-Seng and Li, Qing},
  booktitle={Proc. ACM SIGKDD Int. Conf. Knowl. Discov. Data Min.},
  pages={6491--6501},
  year={2024}
}

@inproceedings{zhu2025collaborative,
  title={Collaborative Retrieval for Large Language Model-based Conversational Recommender Systems},
  author={Zhu, Yaochen and Wan, Chao and Steck, Harald and Liang, Dawen and Feng, Yesu and Kallus, Nathan and Li, Jundong},
  booktitle={Proc. World Wide Web Conf.},
  pages={3323--3334},
  year={2025}
}

@inproceedings{xu2025rallrec,
  title={RALLRec: Improving Retrieval Augmented Large Language Model Recommendation with Representation Learning},
  author={Xu, Jian and Luo, Sichun and Chen, Xiangyu and Huang, Haoming and Hou, Hanxu and Song, Linqi},
  booktitle={Companion Proceedings of the ACM on Web Conference 2025},
  pages={1436--1440},
  year={2025}
}

@inproceedings{li2025g,
  title={G-refer: Graph retrieval-augmented large language model for explainable recommendation},
  author={Li, Yuhan and Zhang, Xinni and Luo, Linhao and Chang, Heng and Ren, Yuxiang and King, Irwin and Li, Jia},
  booktitle={Proc. World Wide Web Conf.},
  pages={240--251},
  year={2025}
}

@techreport{page1999pagerank,
  title={The PageRank citation ranking: Bringing order to the web.},
  author={Page, Lawrence and Brin, Sergey and Motwani, Rajeev and Winograd, Terry},
  year={1999},
  institution={Stanford infolab}
}

@article{kruskal1956shortest,
  title={On the shortest spanning subtree of a graph and the traveling salesman problem},
  author={Kruskal, Joseph B},
  journal={Proceedings of the American Mathematical society},
  volume={7},
  number={1},
  pages={48--50},
  year={1956},
  publisher={JSTOR}
}

@inproceedings{mavromatis2025gnn,
  title={GNN-RAG: Graph neural retrieval for efficient large language model reasoning on knowledge graphs},
  author={Mavromatis, Costas and Karypis, George},
  booktitle={Findings of the Association for Computational Linguistics: ACL 2025},
  pages={16682--16699},
  year={2025}
}

@inproceedings{borgeaud2022improving,
  title={Improving language models by retrieving from trillions of tokens},
  author={Borgeaud, Sebastian and Mensch, Arthur and Hoffmann, Jordan and Cai, Trevor and Rutherford, Eliza and Millican, Katie and Van Den Driessche, George Bm and Lespiau, Jean-Baptiste and Damoc, Bogdan and Clark, Aidan and others},
  booktitle={Proc. Int. Conf. Mach. Learn.},
  pages={2206--2240},
  year={2022},
  organization={PMLR}
}

@inproceedings{jiang2023active,
  title={Active retrieval augmented generation},
  author={Jiang, Zhengbao and Xu, Frank F and Gao, Luyu and Sun, Zhiqing and Liu, Qian and Dwivedi-Yu, Jane and Yang, Yiming and Callan, Jamie and Neubig, Graham},
  booktitle={Proceedings of the 2023 Conference on Empirical Methods in Natural Language Processing},
  pages={7969--7992},
  year={2023}
}

@article{han2024retrieval,
  title={Retrieval-augmented generation with graphs (graphrag)},
  author={Han, Haoyu and Wang, Yu and Shomer, Harry and Guo, Kai and Ding, Jiayuan and Lei, Yongjia and Halappanavar, Mahantesh and Rossi, Ryan A and Mukherjee, Subhabrata and Tang, Xianfeng and others},
  journal={arXiv preprint arXiv:2501.00309},
  year={2024}
}

@inproceedings{liao2024llara,
  title={Llara: Large language-recommendation assistant},
  author={Liao, Jiayi and Li, Sihang and Yang, Zhengyi and Wu, Jiancan and Yuan, Yancheng and Wang, Xiang and He, Xiangnan},
  booktitle={Proceedings of the 47th International ACM SIGIR Conference on Research and Development in Information Retrieval},
  pages={1785--1795},
  year={2024}
}

@inproceedings{zheng2024adapting,
  title={Adapting large language models by integrating collaborative semantics for recommendation},
  author={Zheng, Bowen and Hou, Yupeng and Lu, Hongyu and Chen, Yu and Zhao, Wayne Xin and Chen, Ming and Wen, Ji-Rong},
  booktitle={2024 IEEE 40th International Conference on Data Engineering (ICDE)},
  pages={1435--1448},
  year={2024},
  organization={IEEE}
}

@inproceedings{hou2024large,
  title={Large language models are zero-shot rankers for recommender systems},
  author={Hou, Yupeng and Zhang, Junjie and Lin, Zihan and Lu, Hongyu and Xie, Ruobing and McAuley, Julian and Zhao, Wayne Xin},
  booktitle={European Conference on Information Retrieval},
  pages={364--381},
  year={2024},
  organization={Springer}
}

@inproceedings{lin2024bridging,
  title={Bridging items and language: A transition paradigm for large language model-based recommendation},
  author={Lin, Xinyu and Wang, Wenjie and Li, Yongqi and Feng, Fuli and Ng, See-Kiong and Chua, Tat-Seng},
  booktitle={Proc. ACM SIGKDD Int. Conf. Knowl. Discov. Data Min.},
  pages={1816--1826},
  year={2024}
}

@article{yang2025cold,
  title={Cold-Start Recommendation with Knowledge-Guided Retrieval-Augmented Generation},
  author={Yang, Wooseong and Zhang, Weizhi and Liu, Yuqing and Han, Yuwei and Wang, Yu and Lee, Junhyun and Yu, Philip S},
  journal={arXiv preprint arXiv:2505.20773},
  year={2025}
}

@article{gumaan2025expertrag,
  title={ExpertRAG: Efficient RAG with Mixture of Experts--Optimizing Context Retrieval for Adaptive LLM Responses},
  author={Gumaan, Esmail},
  journal={arXiv preprint arXiv:2504.08744},
  year={2025}
}

@article{maragheh2025arag,
  title={ARAG: Agentic Retrieval Augmented Generation for Personalized Recommendation},
  author={Maragheh, Reza Yousefi and Vadla, Pratheek and Gupta, Priyank and Zhao, Kai and Inan, Aysenur and Yao, Kehui and Xu, Jianpeng and Kanumala, Praveen and Cho, Jason and Kumar, Sushant},
  journal={arXiv preprint arXiv:2506.21931},
  year={2025}
}

@article{gao2023retrieval,
  title={Retrieval-augmented generation for large language models: A survey},
  author={Gao, Yunfan and Xiong, Yun and Gao, Xinyu and Jia, Kangxiang and Pan, Jinliu and Bi, Yuxi and Dai, Yixin and Sun, Jiawei and Wang, Haofen and Wang, Haofen},
  journal={arXiv preprint arXiv:2312.10997},
  volume={2},
  number={1},
  year={2023}
}

@article{lewis2020retrieval,
  title={Retrieval-augmented generation for knowledge-intensive nlp tasks},
  author={Lewis, Patrick and Perez, Ethan and Piktus, Aleksandra and Petroni, Fabio and Karpukhin, Vladimir and Goyal, Naman and K{\"u}ttler, Heinrich and Lewis, Mike and Yih, Wen-tau and Rockt{\"a}schel, Tim and others},
  journal={Proc. Adv. Neural Inf. Process. Syst.},
  volume={33},
  pages={9459--9474},
  year={2020}
}

@article{huang2025towards,
  title={Towards Next-Generation Recommender Systems: A Benchmark for Personalized Recommendation Assistant with LLMs},
  author={Huang, Jiani and Wang, Shijie and Ning, Liang-bo and Fan, Wenqi and Wang, Shuaiqiang and Yin, Dawei and Li, Qing},
  journal={arXiv preprint arXiv:2503.09382},
  year={2025}
}

@inproceedings{ding2023modeling,
  title={Modeling multi-relational connectivity for personalized fashion matching},
  author={Ding, Yujuan and Mok, PY and Bin, Yi and Yang, Xun and Cheng, Zhiyong},
  booktitle={Proceedings of the 31st ACM international conference on multimedia},
  pages={7047--7055},
  year={2023}
}

@article{wang2025graph,
  title={Graph machine learning in the era of large language models (llms)},
  author={Wang, Shijie and Huang, Jiani and Chen, Zhikai and Song, Yu and Tang, Wenzhuo and Mao, Haitao and Fan, Wenqi and Liu, Hui and Liu, Xiaorui and Yin, Dawei and others},
  journal={ACM Transactions on Intelligent Systems and Technology},
  volume={16},
  number={5},
  pages={1--40},
  year={2025},
  publisher={ACM New York, NY}
}

@article{wang2024multi,
  title={Multi-agent attacks for black-box social recommendations},
  author={Wang, Shijie and Fan, Wenqi and Wei, Xiao-Yong and Mei, Xiaowei and Lin, Shanru and Li, Qing},
  journal={ACM Transactions on Information Systems},
  volume={43},
  number={1},
  pages={1--26},
  year={2024},
  publisher={ACM New York, NY}
}

\appendix

\section*{Appendix}

\section{Implementation Details} \label{app:implementation}
The implementation details of \ours{} are summarized in Table~\ref{tab:hyper}. 
The first block reports the optimization-related hyper-parameters for the Mixture-of-Experts retrieval agent and the knowledge alignment agent, including the learning rate ($3\times10^{-4}$), discount factor ($\gamma=0.99$), GAE parameter ($\lambda=0.95$), retrieval budget (Top-$M$ with $M=3$), clipping coefficient ($\epsilon=0.2$), and training epochs.
The optimization is set with batch size 64, buffer size 2048, and gradient clipping with max norm 0.5. The second block lists the preference optimization settings for the recommendation agent, including the temperature $\beta=0.2$ and the number of hard negatives $N=10$.
The third block presents the shared reward hyper-parameters for marginal information gain, namely the MIG weight $\alpha=0.2$ and cost penalty $\eta=0.005$.
The remaining blocks provide the parameter-efficient fine-tuning configuration (LoRA rank $r=16$, scaling $\alpha_{\mathrm{LoRA}}=32$, dropout 0.1, and int8/fp16 settings) and the decoding/evaluation protocol (decoding temperature 0.8, candidate pool size 20, and Recall@$K$ with $K\in\{3,5\}$). In addition, for KG indexing and similarity-based retrieval, we encode queries, entities, and triples into the same semantic space using a pre-trained sentence encoder, \texttt{sentence-transformers/all-MiniLM-L6-v2}. This encoder is used to build the entity and triple vector databases, and retrieval is conducted via Top-$M$ similarity computation over the embedding space.

\section{MMAPO: Detailed Update Rules}\label{app:mmapo_updates}
With advantages from Eq.~\eqref{eq:gae} and the shared reward in Eq.~\eqref{eq:rtotal}, MMAPO updates the three agents in a coordinated manner.
As the retrieval and alignment agents affect the recommendation through sequential decisions, we adopt a clipped update rule to keep each policy update stable and prevent overly large changes between iterations.
Specifically, for the Mixture-of-Experts retrieval agent and the knowledge preference alignment agent, we maximize the following objective:
\begin{equation}
\begin{aligned}
\mathcal{J}_{\mathrm{prox}}(\pi)
&=
\mathbb{E}\!\left[
\min\!\left(\rho_t A_t^{\mathrm{GAE}},\ \mathrm{clip}(\rho_t,1-\epsilon,1+\epsilon)A_t^{\mathrm{GAE}}\right)
\right],
\\
\rho_t
&=\exp\!\left(\log \pi(a_t\mid u_t)-\log \pi_{\mathrm{old}}(a_t\mid u_t)\right),
\label{eq:prox_clip}
\end{aligned}
\end{equation}
where $a_t$ is the action at step $t$ (i.e., selecting an expert or generating a token), $u_t$ denotes the corresponding input to the policy, and $\epsilon$ controls the update range.

For the contrastive learning-reinforced recommendation agent, we optimize the contrastive preference objective in Section~\ref{sec:pref_opt_rec}, which encourages the model to assign a higher probability to the target item than to hard negatives.
By updating the three agents, MMAPO enables \ours{} to learn when to retrieve, how to present KG knowledge, and how to leverage the aligned knowledge for recommendation.
The core procedure of the MMAPO algorithm is presented in Algorithm~\ref{alg:mmapo_core}.

\begin{table}[htbp]
  \centering
  \caption{Hyper-parameter settings of \ours{}.}
  \scalebox{0.82}{
  \begin{tabular}{l|c}
    \toprule
    \textbf{Item} & \textbf{Value} \\
    \midrule
    \multicolumn{2}{c}{\cellcolor{gray!15}\textit{\textbf{Policy Optimization (Retrieval \& Alignment)}}} \\
    learning rate & $3\times 10^{-4}$ \\
    discount factor $\gamma$ & $0.99$ \\
    GAE parameter $\lambda$ & $0.95$ \\
    Retrieval budget $M$ & $3$ \\
    clipping coefficient $\epsilon$ & $0.2$ \\
    epochs & $3$ \\
    \midrule
    \multicolumn{2}{c}{\cellcolor{gray!15}\textit{\textbf{Preference Optimization (Recommendation)}}} \\
    temperature $\beta$ & $0.2$ \\
    number of hard negatives $N$ & $10$ \\
    \midrule
    \multicolumn{2}{c}{\cellcolor{gray!15}\textit{\textbf{Shared Reward (MIG)}}} \\
    MIG weight $\alpha$ & $0.2$ \\
    cost penalty $\eta$ & $0.005$ \\
    \midrule
    \multicolumn{2}{c}{\cellcolor{gray!15}\textit{\textbf{Parameter-efficient Fine-tuning (LoRA)}}} \\
    LoRA rank $r$ & $16$ \\
    LoRA scaling $\alpha_{\mathrm{LoRA}}$ & $32$ \\
    int8 & True \\
    fp16 & True \\
    LoRA dropout & $0.1$ \\
    \midrule
    \multicolumn{2}{c}{\cellcolor{gray!15}\textit{\textbf{Decoding \& Evaluation Protocol}}} \\
    decoding temperature & $0.8$ \\
    candidate pool size & $20$  \\
    evaluated Recall@$K$ & $K\in\{3,5\}$ \\
    \bottomrule
  \end{tabular}}
  \label{tab:hyper}
\end{table}

\begin{algorithm}[htbp]
\caption{\ours{} (MMAPO Training)}
\label{alg:mmapo_core}
\small
\begin{algorithmic}[1]
\REQUIRE Training set $\mathcal{D}$ with instances $(q,\mathcal{O},o^\star)$; KG $G$; retrieval experts $\mathcal{E}_{\mathrm{exp}}$;
retrieval policy $\pi_\theta$ with value $V_\phi$; alignment policy $\pi_\psi$ with value head $V_\psi$; recommender policy $\pi_\omega$ with reference $\pi_{\mathrm{ref}}$;
hyper-parameters $\gamma,\lambda,\epsilon$ (advantage estimation and proximal update), $\alpha,\eta$ (MIG), $\beta$ (preference update), $N$ (hard negatives), retrieval budget $M$.
\STATE Initialize parameters $(\theta,\phi,\psi,\omega)$.
\FOR{each training iteration}
    \STATE Sample a mini-batch $\mathcal{B}\subset\mathcal{D}$.
    \STATE Initialize buffers $\mathcal{T}$ (trajectories) and $\mathcal{P}$ (preference pairs).
    
    \FOR{each $(q,\mathcal{O},o^\star)\in\mathcal{B}$}
        \STATE Construct retrieval state $s$ from $q$.
        \STATE Select retrieval expert $e\sim \pi_\theta(\cdot\mid s)$ and retrieve structured knowledge $\mathcal{K}_e$ from $G$ with budget $M$.
        \STATE Convert $\mathcal{K}_e$ into a stable textual draft $\hat{\mathcal{K}}\leftarrow \mathrm{Temp}(\mathcal{K}_e)$.
        \STATE Generate aligned evidence $\tilde{\mathcal{K}}\sim \pi_\psi(\cdot\mid q,\hat{\mathcal{K}})$.
        \STATE Predict item distribution $P_{\mathrm{expert}}(\cdot)\leftarrow \pi_\omega(\cdot\mid q,\tilde{\mathcal{K}},\mathcal{O})$ and obtain output $o$.
        
        \STATE Compute shared reward $R_{\mathrm{total}} \leftarrow R_{\mathrm{rec}} + \alpha\cdot R_{\mathrm{MIG}}$,
        where $R_{\mathrm{rec}}$ measures recommendation correctness/confidence and
        $R_{\mathrm{MIG}}$ measures marginal information gain with a retrieval cost penalty (using Expert~1 as the baseline to form $P_{\mathrm{base}}$).
        
        \STATE Store the coupled trajectory (states/contexts and actions for expert selection and alignment token generation) into $\mathcal{T}$ with terminal reward $R_{\mathrm{total}}$.
        
        \STATE Mine hard negatives $\mathcal{N}_{\mathrm{hard}}\leftarrow \operatorname{TopN}_{o'\in\mathcal{O}\setminus\{o^\star\}} \pi_\omega(o'\mid q,\tilde{\mathcal{K}},\mathcal{O})$.
        \STATE Add preference pairs $(x,o^\star,o^-)$ into $\mathcal{P}$ for each $o^-\in\mathcal{N}_{\mathrm{hard}}$, where $x=(q,\tilde{\mathcal{K}})$.
    \ENDFOR
    
    \STATE \textbf{Advantage estimation.}
    \STATE Estimate advantages for retrieval using $V_\phi(s)$ and for alignment using $V_\psi(c)$ (Eq.~\ref{eq:gae}).
    
    \STATE \textbf{Coordinated policy update (MMAPO).}
    \STATE Update the retrieval selector $\pi_\theta$ and alignment policy $\pi_\psi$ via a proximal clipped policy update driven by the estimated advantages.
    \STATE Update the recommender $\pi_\omega$ via preference-based contrastive optimization on $\mathcal{P}$ with reference $\pi_{\mathrm{ref}}$.
\ENDFOR
\RETURN Trained parameters $(\theta^\ast,\psi^\ast,\omega^\ast)$.
\end{algorithmic}
\end{algorithm}

\section{Retrieval Complexity}\label{app:complexity}
We analyze the retrieval complexity of \ours{}, compared with two representative baselines.
Table~\ref{tab:retrieval_complexity} summarizes the worst-case retrieval time complexity.
Let $M$ denote the retrieval budget (Top-$M$), $L$ the user history length, and $d$ the embedding dimension. We use $|K|$ to denote the size of returned structured knowledge (e.g., the number of triples/edges).
G-retriever performs query-level retrieval by retrieving relevant nodes and triples/edges from their embedding spaces and then constructing a connected subgraph (e.g., via a Steiner-tree-style procedure), resulting in a cost of $O\!\left((|\mathcal{V}|+|\mathcal{T}|)\cdot(d+\log M) + |K|\right)$.
In contrast, K-RagRec indexes multi-hop item-centered subgraphs in a single vector database and conducts similarity search for each item in the user history. Thus, its retrieval complexity scales linearly with the user history length $L$, i.e., $O\!\left(L\cdot\big(|\mathcal{S}|\cdot(d+\log M) + |K|\big)\right)$.

Our Mixed KG experts span different retrieval granularities with distinct costs.
Expert~1 incurs no retrieval cost.
Expert~2 performs Top-$M$ triple retrieval over $\mathcal{T}$, costing $O\!\left(|\mathcal{T}|\cdot(d+\log M)\right)$.
Expert~3 and Expert~4 first retrieve Top-$M$ seed entities from $\mathcal{V}$ and then expand/construct subgraphs. Therefore, the complexity is $O\!\left(|\mathcal{V}|\cdot(d+\log M) + |K_e|\right)$ for $e\in\{3,4\}$, where $|K_e|$ is linear in the size of the returned subgraph.

Importantly, \ours{} does not commit to a single retrieval strategy.
Instead, it flexibly routes each query to an appropriate expert, enabling a query-dependent trade-off between retrieval cost and knowledge utility.
As a result, \ours{} can avoid unnecessary heavy retrieval for simple queries while still invoking richer subgraph retrieval when needed, yielding a favorable accuracy--efficiency balance compared with baselines that rely on a fixed retrieval pipeline or perform retrieval repeatedly along the user history.

\begin{table}[htbp]
\centering
\caption{Comparison of retrieval time complexity. $M$ is the retrieval budget (Top-$M$) and $L$ is the user history length.}
\label{tab:retrieval_complexity}
\scalebox{0.72}{
\begin{tabular}{l|l}
\toprule
\textbf{Method} & \textbf{Online retrieval time complexity} \\
\midrule
G-retriever
& $O\!\left((|\mathcal{V}|+|\mathcal{T}|)\cdot(d+\log M) + |K|\right)$ \\

K-RagRec 
& $O\!\left(L\cdot\big(|\mathcal{S}|\cdot(d+\log M) + |K|\big)\right)$ \\
\midrule
\textbf{Expert 1 (DirectGenerator)} 
& $O(1)$ \\

\textbf{Expert 2 (TripleRetriever)} 
& $O\!\left(|\mathcal{T}|\cdot(d+\log M)\right)$ \\

\textbf{Expert 3 (SubgraphRetriever)}  
& $O\!\left(|\mathcal{V}|\cdot(d+\log M) + |K_e|\right)$, $e=3$ \\

\textbf{Expert 4 (ConnectedGraphRetriever)} 
& $O\!\left(|\mathcal{V}|\cdot(d+\log M) + |K_e|\right)$, $e=4$ \\

\midrule
\textbf{\ours{}} 
& depends on the selected expert above \\

\bottomrule
\end{tabular}}
\end{table}

\begin{table}[t]
  \centering
  \caption{ Cold-start recommendation performance comparison with the strongest KG-RAG-enhanced LLM recommendation methods on LLaMA3-8B across three metrics. The best performances are labeled in bold. }
    \scalebox{0.95}
{
    \begin{tabular}{c|c|c|c}
    \toprule
    \multirow{1}{*}{\textbf{Methods}}
    & \multicolumn{1}{c|}{\textbf{Accuracy}} & \multicolumn{1}{c|}{\textbf{Recall@3}} & \multicolumn{1}{c}{\textbf{Recall@5}} \\
    \midrule
    LLaMA3-8B\textsubscript{zero-shot}                             & 0.083  &N/A  &N/A \\
    G-retriever &0.356  &0.641 &0.738  \\
    K-RagRec & 0.392 & 0.706 & 0.748 \\
    \rowcolor{yellow!15} \textbf{\ours} &\textbf{0.501}  &\textbf{0.778} &\textbf{0.846}  \\
    \bottomrule
    \end{tabular}%
}
  \label{tab:cold_start}%
\end{table}%

\begin{figure*}[htbp]
\vskip -0.1in
\centering
\subfigure[KG for MovieLens-20M]
{\includegraphics[width=0.98\columnwidth]{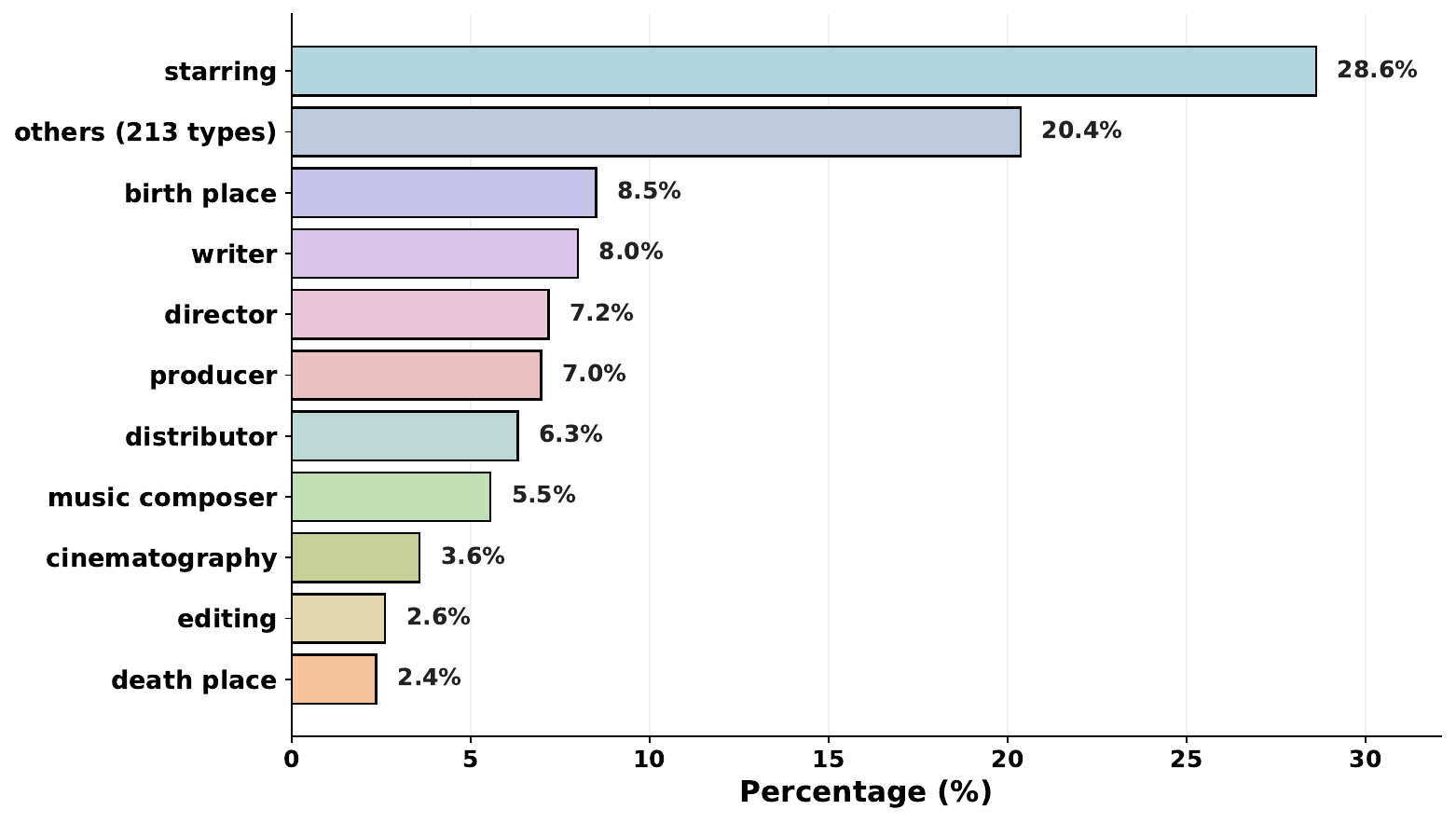}}
\subfigure[KG for LFM-1K]{\includegraphics[width=0.98\columnwidth]{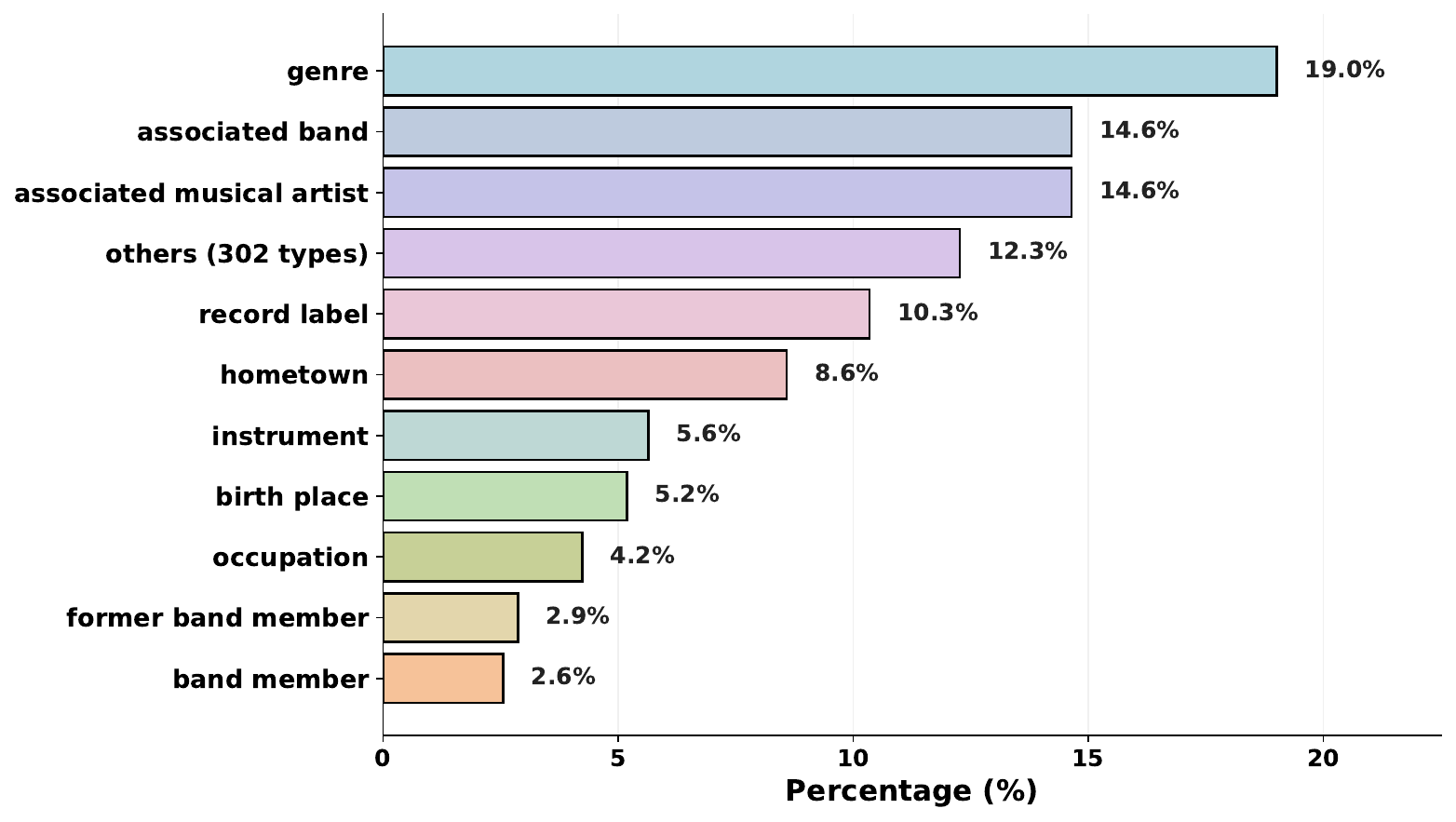}}
\caption{Relation-type distributions of the constructed KGs for (a) MovieLens-20M and (b) LFM-1K.}\label{fig:kg_rel_dist}
\end{figure*}

\begin{figure}[htbp]
    \centering
    \includegraphics[width=0.8\linewidth]{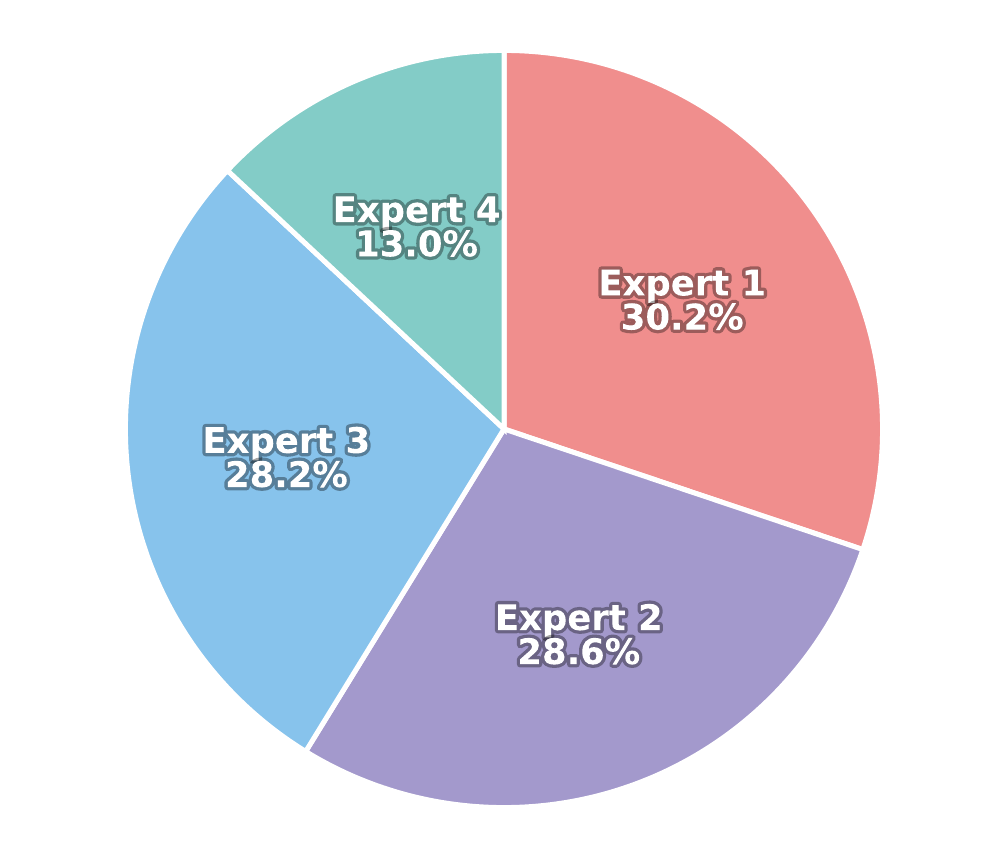}
    \vskip -0.1in
    \caption{Expert selection distribution of the Mixture-of-Experts retrieval agent on LFM-1K with Mistral-7B.  }
    \label{fig:expert_dist}
    \vskip -0.2in
\end{figure}

\section{Expert Selection Distribution}\label{sec:expert_dist}

We further analyze the routing behavior of the Mixture-of-Experts retrieval agent under both standard and cold-start settings.
Figure~\ref{fig:expert_dist} reports the expert selection distribution on LFM-1K with the Mistral-7B backbone.
The agent allocates most instances to Expert~1--3, with proportions of 30.2\%, 28.6\%, and 28.2\%, respectively, while Expert~4 is selected less frequently (13.0\%).
This pattern indicates that, for LFM-1K, the selector predominantly relies on lightweight or moderate-granularity retrieval and invokes the most expensive expert only for a smaller subset of instances.

\begin{figure}[htbp]
    \centering
    \includegraphics[width=0.8\linewidth]{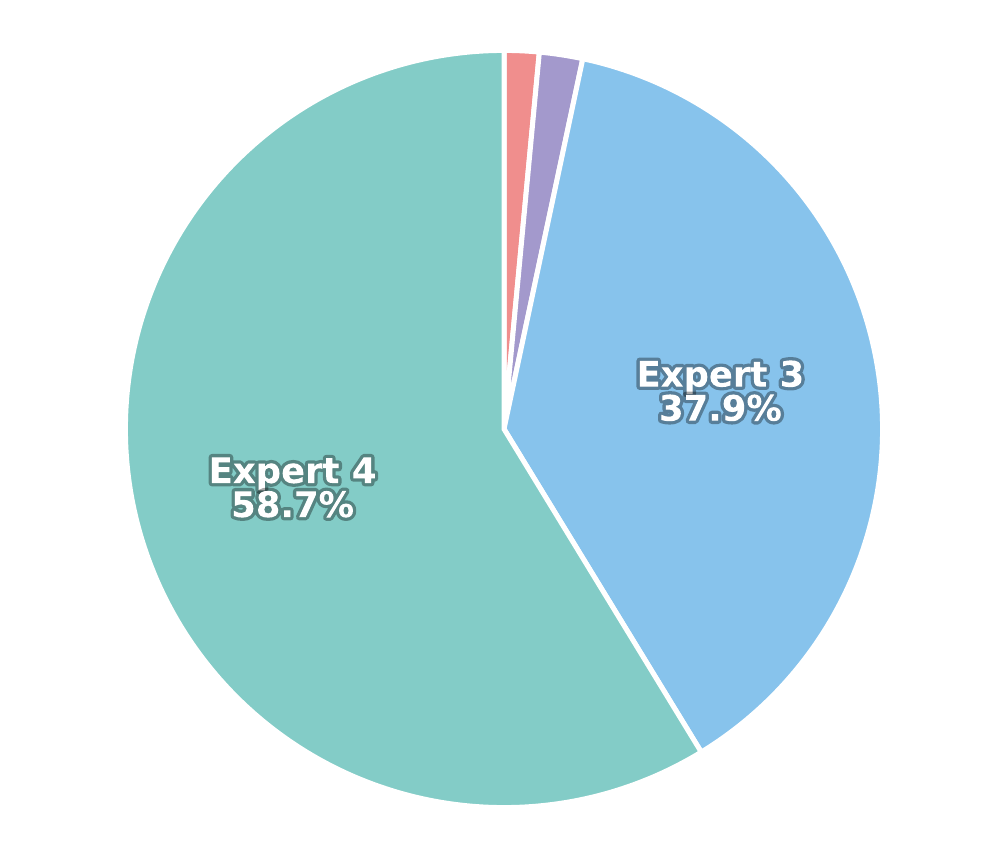}
    \vskip -0.1in
    \caption{Expert selection distribution of the Mixture-of-Experts retrieval agent on cold-start recommendation scenario.  }
    \label{fig:expert_dist_cold}
    \vskip -0.2in
\end{figure}

We also examine the selection distribution in the cold-start scenario.
As shown in Figure~\ref{fig:expert_dist_cold}, \ours{} chooses Expert~3 for 37.9\% of instances and Expert~4 for 58.7\% of instances, suggesting that cold-start queries more often benefit from higher-granularity relational knowledge.
These results highlight that MixRAGRec adapts retrieval granularity to different data contexts rather than using a fixed strategy.

\section{Cold-start Recommendation Study}
Cold-start recommendation remains an important challenge in practice.
To evaluate \ours{} under this setting, we compare it with strong KG-RAG-enhanced baselines on MovieLens-1M using the LLaMA3-8B backbone, as reported in Table~\ref{tab:cold_start}.
Zero-shot prompting performs poorly in the cold-start scenario, which further motivates grounding recommendations with RAG.
Compared with other KG-RAG baselines, \ours{} achieves the best results and remains robust under the cold-start setting, demonstrating the effectiveness of our framework.

We further examine the expert-selection behavior in cold-start scenarios.
Figure~\ref{fig:expert_dist_cold} shows that \ours{} selects subgraph-level retrieval (Expert~3) for 37.9\% of queries and connected-graph retrieval (Expert~4) for 58.7\% of queries, suggesting that cold-start instances frequently require richer relational context.
This trend aligns with the gains in Table~\ref{tab:cold_start} and highlights the effectiveness of \ours{} in adapting retrieval granularity to the cold-start recommendations.

\section{More Related Work}\label{app:related}
Large language models (LLMs) have recently emerged as a promising paradigm for recommendation due to their strong semantic understanding and generalization capabilities. 
Early research reformulates recommendations as a language modeling task to leverage LLMs’ language understanding and in-context learning capabilities for user preference modeling~\cite{geng2022recommendation,bao2023tallrec,lin2024bridging}. 
For example, LLMRank~\cite{hou2024large} employs instruction-based prompts constructed from historical interactions to perform item re-ranking, whereas TALLRec~\cite{bao2023tallrec} introduces an instruction-tuning framework that directly recommends target items by modeling user interest from text-enriched historical interactions (e.g., item titles). 
To further enhance performance, subsequent efforts incorporate collaborative signals into LLM-based recommendations. Specifically, LLaRa~\cite{liao2024llara} and CoLLM~\cite{zhang2025collm} treat collaborative information as an additional modality and align it with the LLM token space through learned mappings.
Moreover, tokenization-based approaches encode external signals into item identifiers, representing items as discrete semantic or collaborative indices to facilitate LLM-based recommendation~\cite{zheng2024adapting,qu2025tokenrec}.
Despite these advancements, LLM-based recommender systems remain constrained by static parametric knowledge and have limitations in obtaining up-to-date information.
Therefore, to avoid costly fine-tuning, it is important to introduce RAG to incorporate up-to-date external knowledge.

\section{Knowledge Graph Statistics}\label{app:kg_stats}
To support the KG-RAG recommendations, we construct an item-centric knowledge graph (KG) for each dataset.
Table~\ref{tab:datasets} reports basic statistics of the three real-world datasets and their associated KGs, including the numbers of users, items, interactions, and the aligned items that appear in both the dataset and the KG (\emph{Items in KG}). Figure~\ref{fig:kg_rel_dist} further visualizes the relation-type distributions for the MovieLens-20M and LFM-1K KGs.
Overall, the KG coverage is substantial across datasets, providing rich, structured knowledge for grounding LLM-based recommendations.
In addition, the KGs exhibit diverse schema complexity, with hundreds of relation types and up to hundreds of thousands of triples, which motivates retrieval designs that can flexibly control the granularity and amount of retrieved knowledge.

\section{Prompts used in \ours{}}
We present the prompts for our \ours{} in Table~\ref{tab:prompt_align} and ~\ref{tab:prompt_rec}. Specifically, the prompt used for Knowledge Preference Alignment Agent is shown in Tables~\ref{tab:prompt_align}, and the prompt used for Contrastive Learning-reinforced Recommendation Agent is shown in Tables~\ref{tab:prompt_rec}.

\section{Case Study}
To provide an intuitive understanding of how \ours{} operates in practice, we present three representative instances from MovieLens-1M and LFM-1K in Tables~\ref{tab:case_ml1m_2}, \ref{tab:case_ml1m}, and \ref{tab:case_lfm1k}.
All cases follow the same pipeline: \ours{} first routes the input instance to a retrieval expert, optionally retrieves structured KG knowledge, aligns the knowledge into concise natural language, and finally produces a recommendation over the candidate pool.

Table~\ref{tab:case_ml1m_2} shows a MovieLens-1M example routed to \textbf{Expert~2 (TripleRetriever)}.
The retrieved triples highlight significant relations related to the user’s recent interests, and the alignment agent converts them into a short, fluent knowledge snippet directly consumable by the recommendation agent.
Conditioned on this aligned knowledge, the model assigns a highly peaked probability mass to the correct option, illustrating how lightweight triple-level retrieval can be sufficient when a small number of key facts already distinguish the target item.

Table~\ref{tab:case_ml1m} presents a MovieLens-1M instance where the selector chooses \textbf{Expert~3 (SubgraphRetriever)}.
Compared with isolated triples, subgraph-level retrieval provides richer relational context.
In this example, the retrieved subgraph captures a compact family-oriented cluster, and the alignment agent converts it into an LLM-friendly description for downstream recommendation.
Conditioned on the aligned knowledge, the recommendation agent ranks the ground-truth option highest, illustrating the benefit of subgraph-level evidence when a multi-relation context is needed.

Finally, Table~\ref{tab:case_lfm1k} shows an LFM-1K instance routed to \textbf{Expert~1 (DirectGenerator)}, where \ours{} skips retrieval and alignment entirely.
The listening history already provides strong textual cues for the backbone model, and bypassing retrieval avoids unnecessary overhead while still yielding a confident correct prediction.
This case highlights that \ours{} can preserve efficiency by not invoking retrieval when external evidence is unlikely to offer additional benefit.

Overall, these cases reflect the key advantage of \ours{} in controlling the effectiveness--utility trade-off.
It can rely on lightweight evidence when it is sufficient (Table~\ref{tab:case_ml1m_2}), avoid over-retrieval that may introduce noise (Table~\ref{tab:case_ml1m}), and skip retrieval altogether for easy instances (Table~\ref{tab:case_lfm1k}).

\begin{table*}[t]
\caption{General prompt template for the Knowledge Preference Alignment Agent.}
\label{tab:prompt_align}
\centering
\begin{minipage}{0.92\textwidth}
\begin{tcolorbox}[
  title=Knowledge Preference Alignment Agent Prompt,
  colback=white,
  colframe=black!60,
  boxrule=0.6pt,
  arc=2pt,
  left=6pt,right=6pt,top=6pt,bottom=6pt,
  fonttitle=\bfseries
]

\tcbsubtitle{Shared Input}
\textbf{User Query:} Watching history: \textit{``Movie A'', ``Movie B'', \ldots, ``Movie J''}.\\
\textbf{User Preferences:} general user\\
\textbf{Domain:} Movie/Music\\
\textbf{History Label:} "watching history"/ "listening history"\\
\textbf{Item Type:} "movie"/"artist"

\par\smallskip\hrule\smallskip

\tcbsubtitle{Expert 2 (Triple-level Retrieval)}
\textbf{Task Description:} Convert the following factual triples into natural language knowledge suitable for movie recommendations.\\
\textbf{Retrieved Triples:} \texttt{\{structured\_triples\_from\_KG\}}\\
\textbf{Instructions:}
\begin{enumerate}\setlength{\itemsep}{0pt}\setlength{\parskip}{0pt}
  \item Extract relevant facts from the triples
  \item Convert to natural fluent text
  \item Focus on genres, themes, directors, and user preferences
\end{enumerate}

\par\smallskip\hrule\smallskip

\tcbsubtitle{Expert 3 (Neighborhood Subgraph)}
\textbf{Task Description:} Transform the following knowledge graph subgraph into natural language suitable for movie recommendations.\\
\textbf{Retrieved Subgraph:} \texttt{\{structured\_subgraph\_from\_KG\}}\\
\textbf{Instructions:}
\begin{enumerate}\setlength{\itemsep}{0pt}\setlength{\parskip}{0pt}
  \item Identify key entities and relationships
\item Describe connection patterns
\item Highlight genres, themes, and stylistic similarities
\item Keep description concise and relevant
\end{enumerate}

\par\smallskip\hrule\smallskip

\tcbsubtitle{Expert 4 (Connected Subgraph)}
\textbf{Task Description:} Transform the following connected knowledge graph into natural language for movie recommendations.\\
\textbf{Retrieved Connected Graph:} \texttt{\{pagerank\_mst\_graph\_from\_KG\}}\\
\textbf{Instructions:}
\begin{enumerate}\setlength{\itemsep}{0pt}\setlength{\parskip}{0pt}
\item Extract core themes from connected entities
\item Describe shared characteristics
\item Focus on genres, directors, and cinematography
\item Provide a coherent narrative summary
\end{enumerate}

\par\smallskip\hrule\smallskip

\tcbsubtitle{Output}
\textbf{Knowledge Summary:} \texttt{\{aligned\_knowledge\_summary\}}

\end{tcolorbox}
\end{minipage}
\end{table*}

\begin{table*}[t]
\caption{General prompt template for the Contrastive Learning-reinforced Recommendation Agent.}
\label{tab:prompt_rec}
\centering
\begin{minipage}{0.92\textwidth}
\begin{tcolorbox}[
  title=Contrastive Learning-reinforced Recommendation Agent Prompt,
  colback=white,
  colframe=black!60,
  boxrule=0.6pt,
  arc=2pt,
  left=6pt,right=6pt,top=6pt,bottom=6pt,
  fonttitle=\bfseries
]
\tcbsubtitle{Shared Input}
\textbf{Domain:} Movie/Music\\
\textbf{History Label:} "watching history"/ "listening history"\\
\textbf{Item Type:} "movie"/"artist"
\par\smallskip\hrule\smallskip

\tcbsubtitle{Default (with KG knowledge)}
\textbf{Instruction:} Based on the user's watching history, select the best movie from options A-T.\\
\textbf{User Query:} Watching history: \textit{``Movie A'', ``Movie B'', \ldots, ``Movie J''}.\\
\textbf{Knowledge:} \{aligned\_knowledge\_summary\}\\

\textbf{Options}: \textit{``A: Movie'', ``B: Movie'', \ldots, ``T: Movie''}.\\
Select a movie from options A to T that the user is most likely to be interested in.

\par\smallskip\hrule\smallskip

\tcbsubtitle{For Expert 0 (no KG knowledge)}
\textbf{Instruction:} Based on the user's watching history, select the best movie from options A-T.\\
\textbf{User Query:} Watching history: \textit{``Movie A'', ``Movie B'', \ldots, ``Movie J''}.\\

\textbf{Options}: \textit{``A: Movie'', ``B: Movie'', \ldots, ``T: Movie''}.\\
Select a movie from options A to T that the user is most likely to be interested in.
\end{tcolorbox}
\end{minipage}
\end{table*}

\begin{table*}[t]
\caption{Case study on Movie with Expert 2 (TripleRetriever).}
\label{tab:case_ml1m_2}
\centering
\begin{minipage}{0.92\textwidth}
\begin{tcolorbox}[
  colback=white,
  colframe=black!60,
  colbacktitle=orange!10,
  coltitle=black,
  boxrule=0.6pt,
  arc=2pt,
  left=6pt,right=6pt,top=6pt,bottom=6pt,
  title=\textbf{Case Study: Movie (TripleRetriever)},
  fonttitle=\bfseries
]
\small

\textbf{(1) Input instance.}\par
\textbf{User Query (watching history):}
\textit{``Parenthood'', ``Dark City'', ``Reservoir Dogs'', ``Aliens'', ``The Sixth Sense'', ``The Terminator'', ``A Simple Plan'', ``Die Hard'', ``Seven'', ``The Long Kiss Goodnight''.}\par
\textbf{Candidate options (A--T):}
\textit{A: Hellbound: Hellraiser II, B: Bring It On, C: Escape from L.A., D: One Tough Cop, E: Boxing Helena, F: Separation, The, G: From the Journals of Jean Seberg, H: Terminator 2: Judgment Day, I: Following, J: The Big Country, K: One Flew Over the Cuckoo's Nest, L: Grandview, U.S.A., M: Cotton Mary, N: Smiling Fish and Goat on Fire, O: School Daze, P: Soldier's Daughter Never Cries, A, Q: Weekend at Bernie's, R: Love and Death on Long Island, S: Roseanna's Grave, T: Maurice.}\par
\textbf{Ground-truth:} \textbf{H}.

\par\smallskip\hrule\smallskip

\textbf{(2) Mixture-of-Experts routing.}\par
\textbf{Selected expert:} \textbf{Expert 2 (TripleRetriever)}.

\par\smallskip\hrule\smallskip

\textbf{(3) KG retrieval output (structured).}\par
\textbf{Retrieved triples:}\par
\begin{itemize}
  \setlength{\itemsep}{2pt}
  \setlength{\topsep}{2pt}
  \setlength{\leftmargin}{14pt}
  \item The Terminator $\xrightarrow{\text{director}}$ James Cameron
  \item Terminator 2: Judgment Day $\xrightarrow{\text{is}}$ highest-grossing film list 
  \item Indiana Jones and the Temple of Doom$\xrightarrow{\text{star}}$ Kate Capshaw 
  \item \ldots
\end{itemize}

\par\smallskip\hrule\smallskip

\textbf{(4) Knowledge Preference Alignment (Agent).}\par
\textbf{Alignment prompt (abridged):}
Convert the following factual triples into natural language knowledge suitable for movie recommendations.\par
\textbf{Aligned knowledge (output):}\par
\emph{Terminator 2: Judgment Day is a movie highly relevant to user queries, as its lead actor is associated with the query and it is related to other related films such as The Terminator. The Terminator, a movie starring Arnold Schwarzenegger, is a highly relevant film related to the user's query \ldots}

\par\smallskip\hrule\smallskip

\textbf{(5) Recommendation (Contrastive Learning-reinforced Rec Agent).}\par
\textbf{Output:} \textbf{H. Terminator 2: Judgment Day}.\par
\textbf{Model confidence / option distribution (top):}
\begin{itemize}
  \setlength{\itemsep}{2pt}
  \setlength{\topsep}{2pt}
  \setlength{\leftmargin}{14pt}
  \item H (Terminator 2): 0.92 \quad $\Leftarrow$ \quad \textbf{(ground truth)}
  \item J (The Big Country): 0.01
  \item M (Cotton Mary): 0.01
  \item N (Smiling Fish and Goat on Fire): 0.01
\end{itemize}

\end{tcolorbox}
\end{minipage}
\end{table*}

\begin{table*}[t]
\caption{Case study on Movie with Expert 3 (SubgraphRetriever).}
\label{tab:case_ml1m}
\centering
\begin{minipage}{0.92\textwidth}
\begin{tcolorbox}[
  colback=white,
  colframe=black!60,
  colbacktitle=orange!10,
  coltitle=black,
  boxrule=0.6pt,
  arc=2pt,
  left=6pt,right=6pt,top=6pt,bottom=6pt,
  title=\textbf{Case Study: Movie (SubgraphRetriever)},
  fonttitle=\bfseries
]
\small

\textbf{(1) Input instance.}\par
\textbf{User Query (watching history):}
\textit{``The Lion King'', ``The Road to El Dorado'', ``Oliver \& Company'', ``The Incredible Journey'', ``Home Alone'', ``Jumanji'', ``Flubber'', ``Anastasia'', ``Labyrinth'', ``Tom and Huck''.}\par
\textbf{Candidate options (A--T):}
\textit{A: Brassed Off, B: Mighty Ducks, The, C: Porky's II: The Next Day, D: Ninotchka, E: The City, F: The Grass Harp, G: Like Water for Chocolate, H: Night of the Comet, I: Shaft, J: The Kid, K: Leave It to Beaver, L: Inspector Gadget, M: Fright Night Part II, N: Chopping Mall, O: Henry: Portrait of a Serial Killer, Part 2, P: Austin Powers: International Man of Mystery, Q: Romancing the Stone, R: NeverEnding Story II: The Next Chapter, S: Men in Black, T: Re-Animator.}\par
\textbf{Ground-truth:} \textbf{L}.

\par\smallskip\hrule\smallskip

\textbf{(2) Mixture-of-Experts routing.}\par
\textbf{Selected expert:} \textbf{Expert 3 (SubgraphRetriever)}.

\par\smallskip\hrule\smallskip

\textbf{(3) KG retrieval output (structured).}\par
\textbf{Retrieved subgraph:}\par
\textbf{Entities:} \{Flubber (Movie), Jumanji (Movie), Oliver \& Company (Movie)\}.\par
\textbf{Relations:}
\begin{itemize}
  \setlength{\itemsep}{2pt}
  \setlength{\topsep}{2pt}
  \setlength{\leftmargin}{14pt}
  \item Flubber $\xrightarrow{\text{genre}}$ family comedy
  \item Flubber $\xrightarrow{\text{theme}}$ playful invention / slapstick
  \item Jumanji $\xrightarrow{\text{genre}}$ family adventure
  \item Jumanji $\xrightarrow{\text{theme}}$ fantasy adventure / kid-friendly
  \item Oliver \& Company $\xrightarrow{\text{genre}}$ animated family film
  \item Oliver \& Company $\xrightarrow{\text{theme}}$ lighthearted, for all ages

  \item \ldots
\end{itemize}

\par\smallskip\hrule\smallskip

\textbf{(4) Knowledge Preference Alignment (Agent).}\par
\textbf{Alignment prompt (abridged):}
Transform the following knowledge graph subgraph into natural language suitable for movie recommendations.\par
\textbf{Aligned knowledge (output):}\par
\emph{The subgraph contains 3 core entities and a compact set of relations, where \textit{Flubber} receives the highest relevance score and serves as the central node. The subgraph connects these entities through recurring family-oriented signals and lightweight entertainment attributes: For \textit{Flubber}, the evidence emphasizes a comedy-focused profile with playful, kid-friendly cues; for \textit{Jumanji}, the relations highlight family adventure elements and a fantasy-leaning storyline \ldots}

\par\smallskip\hrule\smallskip

\textbf{(5) Recommendation (Contrastive Learning-reinforced Rec Agent).}\par
\textbf{Output:} \textbf{L Inspector Gadget}.\par
\textbf{Model confidence / option distribution (top):}
\begin{itemize}
  \setlength{\itemsep}{2pt}
  \setlength{\topsep}{2pt}
  \setlength{\leftmargin}{14pt} 
  \item L (Inspector Gadget): 0.55 \quad $\Leftarrow$ \quad \textbf{(ground truth)}
  \item B (Mighty Ducks): 0.22
  \item S (Men in Black): 0.09
  \item J (The Kid): 0.05
\end{itemize}

\end{tcolorbox}
\end{minipage}
\end{table*}

\begin{table*}[t]
\caption{Case study on Music with Expert 1 (DirectGenerator).}
\label{tab:case_lfm1k}
\centering
\begin{minipage}{0.92\textwidth}
\begin{tcolorbox}[
  colback=white,
  colframe=black!60,
  colbacktitle=orange!10,
  coltitle=black,
  boxrule=0.6pt,
  arc=2pt,
  left=6pt,right=6pt,top=6pt,bottom=6pt,
  title=\textbf{Case Study: Music (DirectGenerator)},
  fonttitle=\bfseries
]
\small

\textbf{(1) Input instance.}\par
\textbf{User Query (listening history):}
\textit{``Boards Of Canada'', ``Zazen Boys'', ``Recloose'', ``Flying Lotus'', ``The Ananda Project'', ``Afterlife'', ``Onra \& Quetzal'', ``Ian O'Brien'', ``Onra \& Quetzal'', ``Recloose''.}\par
\textbf{Candidate options (A--T):}
\textit{A: Dumptruck, B: Dj Lol, C: The Livids, D: Hei, E: Tactics, F: Onra \& Quetzal, G: Eric Smax \& Terri B., H: Dan Curtin, I: Pebbles, J: Jochen Trappe, K: Versbox, L: Hans Zimmer, M: Melanie Doane, N: Ana D, O: Dave The Hustler, P: Jo O'Meara, Q: Codes In The Clouds, R: Fatal Bazooka, S: Only Fools And Horses, T: Pell Mell.}\par
\textbf{Ground-truth:} \textbf{F}.

\par\smallskip\hrule\smallskip

\textbf{(2) Mixture-of-Experts routing.}\par
\textbf{Selected expert:} \textbf{Expert 1 (DirectGenerator)}.

\par\smallskip\hrule\smallskip

\textbf{(3) KG retrieval output (structured).}\par
\textbf{Retrieved knowledge:} \emph{Skipped} (DirectGenerator uses no KG retrieval).

\par\smallskip\hrule\smallskip

\textbf{(4) Knowledge Preference Alignment (Agent).}\par
\textbf{Aligned knowledge (output):} \emph{Skipped} (no retrieved knowledge; $\tilde{\mathcal{K}}=\varnothing$).

\par\smallskip\hrule\smallskip

\textbf{(5) Recommendation (Contrastive Learning-reinforced Rec Agent).}\par
\textbf{Output:} \textbf{F Onra \& Quetzal}.\par
\textbf{Model confidence / option distribution (top):}
\begin{itemize}
  \setlength{\itemsep}{2pt}
  \setlength{\topsep}{2pt}
  \setlength{\leftmargin}{14pt}
  \item F (Onra \& Quetzal): 0.76 \quad $\Leftarrow$ \quad \textbf{(ground truth)}
  \item L (Hans Zimmer): 0.12
  \item P (Jo O'Meara): 0.06
  \item K (Versbox): 0.03
\end{itemize}

\end{tcolorbox}
\end{minipage}
\end{table*}

\end{document}